\documentclass[aps,prc,twocolumn,superscriptaddress,showpacs,showkeys]{revtex4-1}

\usepackage{amsmath}
\usepackage{graphicx}
\usepackage{wasysym}

\begin{document}


\title{$\beta$-decay half-lives and $\beta$-delayed neutron emission probabilities for several isotopes of Au, Hg, Tl, Pb and Bi, beyond N=126}


	\author{R.~Caballero-Folch}
	\email[Author contact: ]{rcaballero-folch@triumf.ca/roger@baeturia.com}
	\affiliation{INTE, DFEN - Universitat Polit\`ecnica de Catalunya, E-08028 Barcelona, Spain}
	\affiliation{TRIUMF, Vancouver, British Columbia V6T 2A3, Canada}
	\author{C.~Domingo-Pardo}
	\affiliation{IFIC, CSIC - Universitat de Val\`encia, E-46071 Val\`encia, Spain}
	\author{J.~Agramunt} 
	\affiliation{IFIC, CSIC - Universitat de Val\`encia, E-46071 Val\`encia, Spain}
	\author{A.~Algora} 
	\affiliation{IFIC, CSIC - Universitat de Val\`encia, E-46071 Val\`encia, Spain}
	\affiliation{Institute of Nuclear Research of the Hungarian Academy of Sciences, Debrecen H-4001, Hungary}
	\author{F.~Ameil}
	\affiliation{GSI Helmholtzzentrum f\"ur Schwerionenforschung GmbH, D-64291 Darmstadt, Germany}
	\author{Y.~Ayyad} 
	\affiliation{Universidade de Santiago de Compostela, E-15782 Santiago de Compostela, Spain}
	\author{J.~Benlliure} 
	\affiliation{Universidade de Santiago de Compostela, E-15782 Santiago de Compostela, Spain}
	\author{M.~Bowry} 
	\affiliation{Department of Physics, University of Surrey, Guildford GU2 7XH, United Kingdom}
	\author{F.~Calvi\~no} 
	\affiliation{INTE, DFEN - Universitat Polit\`ecnica de Catalunya, E-08028 Barcelona, Spain}
	\author{D.~Cano-Ott} 
	\affiliation{CIEMAT, E-28040 Madrid, Spain}
	\author{G.~Cort\`es} 
	\affiliation{INTE, DFEN - Universitat Polit\`ecnica de Catalunya, E-08028 Barcelona, Spain}
	\author{T.~Davinson} 
	\affiliation{University of Edinburgh, Edinburgh EH9 3JZ, United Kingdom}
	\author{I.~Dillmann}
	\affiliation{TRIUMF, Vancouver, British Columbia V6T 2A3, Canada}
	\affiliation{GSI Helmholtzzentrum f\"ur Schwerionenforschung GmbH, D-64291 Darmstadt, Germany}
	\affiliation{Justus-Liebig-Universit\"at Giessen, D-35392 Giessen, Germany}
	\author{A.~Estrade}
	\affiliation{GSI Helmholtzzentrum f\"ur Schwerionenforschung GmbH, D-64291 Darmstadt, Germany}
	\affiliation{St. Mary's University, Halifax, Nova Scotia B3H 3C3, Canada}
	\author{A.~Evdokimov}
	\affiliation{GSI Helmholtzzentrum f\"ur Schwerionenforschung GmbH, D-64291 Darmstadt, Germany}
	\affiliation{Justus-Liebig-Universit\"at Giessen, D-35392 Giessen, Germany}
	\author{T.~Faestermann}
	\affiliation{Physik Department E12, Technische Universit\"at M\"unchen, D-85748 Garching, Germany}
	\author{F.~Farinon}
	\affiliation{GSI Helmholtzzentrum f\"ur Schwerionenforschung GmbH, D-64291 Darmstadt, Germany}
	\author{D.~Galaviz}
	\affiliation{Centro de F\'isica Nuclear da Universidade de Lisboa, 169-003 Lisboa, Portugal}
	\author{A.R.~Garc\'ia}
	\affiliation{CIEMAT, E-28040 Madrid, Spain}
	\author{H.~Geissel}
	\affiliation{GSI Helmholtzzentrum f\"ur Schwerionenforschung GmbH, D-64291 Darmstadt, Germany}
	\affiliation{Justus-Liebig-Universit\"at Giessen, D-35392 Giessen, Germany}
	\author{W.~Gelletly}
	\affiliation{Department of Physics, University of Surrey, Guildford GU2 7XH, United Kingdom}
	\author{R.~Gernh\"auser}
	\affiliation{Physik Department E12, Technische Universit\"at M\"unchen, D-85748 Garching, Germany}
	\author{M.B.~G\'omez-Hornillos}
	\affiliation{INTE, DFEN - Universitat Polit\`ecnica de Catalunya, E-08028 Barcelona, Spain}
	\author{C.~Guerrero}
	\affiliation{CERN Physics Department, CH-1211 Geneve, Switzerland}
	\affiliation{Universidad de Sevilla, E-41080 Sevilla, Spain}
	\author{M.~Heil}
	\affiliation{GSI Helmholtzzentrum f\"ur Schwerionenforschung GmbH, D-64291 Darmstadt, Germany}
	\author{C.~Hinke}
	\affiliation{Physik Department E12, Technische Universit\"at M\"unchen, D-85748 Garching, Germany}
	\author{R.~Kn\"obel}
	\affiliation{GSI Helmholtzzentrum f\"ur Schwerionenforschung GmbH, D-64291 Darmstadt, Germany}
	\author{I.~Kojouharov}
	\affiliation{GSI Helmholtzzentrum f\"ur Schwerionenforschung GmbH, D-64291 Darmstadt, Germany}
	\author{J.~Kurcewicz}
	\affiliation{GSI Helmholtzzentrum f\"ur Schwerionenforschung GmbH, D-64291 Darmstadt, Germany}
	\author{N.~Kurz}
	\affiliation{GSI Helmholtzzentrum f\"ur Schwerionenforschung GmbH, D-64291 Darmstadt, Germany}
	\author{Yu.A.~Litvinov}
	\affiliation{GSI Helmholtzzentrum f\"ur Schwerionenforschung GmbH, D-64291 Darmstadt, Germany}
	\author{L.~Maier}
	\affiliation{Physik Department E12, Technische Universit\"at M\"unchen, D-85748 Garching, Germany}
	\author{J.~Marganiec}
	\affiliation{ExtreMe Mater Institute, D-64291 Darmstadt, Germany}
	\author{M.~Marta}
	\affiliation{GSI Helmholtzzentrum f\"ur Schwerionenforschung GmbH, D-64291 Darmstadt, Germany}
	\affiliation{Justus-Liebig-Universit\"at Giessen, D-35392 Giessen, Germany}
	\author{T.~Mart\'inez}
	\affiliation{CIEMAT, E-28040 Madrid, Spain}
	\author{F.~Montes}
	\affiliation{NSCL, Michigan State University, East Lansing, MI 48824, USA}
	\affiliation{Joint Institute for Nuclear Astrophysics, Notre Dame, IN 46615, USA}
	\author{I.~Mukha}
	\affiliation{GSI Helmholtzzentrum f\"ur Schwerionenforschung GmbH, D-64291 Darmstadt, Germany}
	\author{D.R.~Napoli}
	\affiliation{Instituto Nazionale di Fisica Nucleare, Laboratori Nazionale di Legnaro, I-35020 Legnaro, Italy}
	\author{C.~Nociforo}
	\affiliation{GSI Helmholtzzentrum f\"ur Schwerionenforschung GmbH, D-64291 Darmstadt, Germany}
	\author{C.~Paradela}
	\affiliation{Universidade de Santiago de Compostela, E-15782 Santiago de Compostela, Spain}
	\author{S.~Pietri}
	\affiliation{GSI Helmholtzzentrum f\"ur Schwerionenforschung GmbH, D-64291 Darmstadt, Germany}
	\author{Zs.~Podoly\'ak}
	\affiliation{Department of Physics, University of Surrey, Guildford GU2 7XH, United Kingdom}
	\author{A.~Prochazka}
	\affiliation{GSI Helmholtzzentrum f\"ur Schwerionenforschung GmbH, D-64291 Darmstadt, Germany}
	\author{S.~Rice}
	\affiliation{Department of Physics, University of Surrey, Guildford GU2 7XH, United Kingdom}
	\author{A.~Riego}
	\affiliation{INTE, DFEN - Universitat Polit\`ecnica de Catalunya, E-08028 Barcelona, Spain}
	\author{B.~Rubio}
	\affiliation{IFIC, CSIC - Universitat de Val\`encia, E-46071 Val\`encia, Spain}
	\author{H.~Schaffner}
	\affiliation{GSI Helmholtzzentrum f\"ur Schwerionenforschung GmbH, D-64291 Darmstadt, Germany}
	\author{Ch.~Scheidenberger}
	\affiliation{GSI Helmholtzzentrum f\"ur Schwerionenforschung GmbH, D-64291 Darmstadt, Germany}
	\affiliation{Justus-Liebig-Universit\"at Giessen, D-35392 Giessen, Germany}
	\author{K.~Smith}
	\affiliation{GSI Helmholtzzentrum f\"ur Schwerionenforschung GmbH, D-64291 Darmstadt, Germany}
	\affiliation{NSCL, Michigan State University, East Lansing, MI 48824, USA}
	\affiliation{Joint Institute for Nuclear Astrophysics, Notre Dame, IN 46615, USA}
	\affiliation{University of Notre Dame, South Bend, IN 46556, USA}
	\affiliation{University of Tennessee, Knoxville, TN 37996, USA}
	\author{E.~Sokol}
	\affiliation{Flerov Laboratory, Joint Institute for Nuclear Research, 141980 Dubna, Russia}
	\author{K.~Steiger}
	\affiliation{Physik Department E12, Technische Universit\"at M\"unchen, D-85748 Garching, Germany}
	\author{B.~Sun}
	\affiliation{GSI Helmholtzzentrum f\"ur Schwerionenforschung GmbH, D-64291 Darmstadt, Germany}
	\author{J.L.~Ta\'in} 
	\affiliation{IFIC, CSIC - Universitat de Val\`encia, E-46071 Val\`encia, Spain}
	\author{M.~Takechi}
	\affiliation{GSI Helmholtzzentrum f\"ur Schwerionenforschung GmbH, D-64291 Darmstadt, Germany}
	\author{D.~Testov}
	\affiliation{Flerov Laboratory, Joint Institute for Nuclear Research, 141980 Dubna, Russia}
	\affiliation{Institute de Physique Nucl\'eaire d'Orsay, F-91405 Orsay, France}
	\author{H.~Weick}
	\affiliation{GSI Helmholtzzentrum f\"ur Schwerionenforschung GmbH, D-64291 Darmstadt, Germany}
	\author{E.~Wilson}
	\affiliation{Department of Physics, University of Surrey, Guildford GU2 7XH, United Kingdom}
	\author{J.S.~Winfield}
	\affiliation{GSI Helmholtzzentrum f\"ur Schwerionenforschung GmbH, D-64291 Darmstadt, Germany}
	\author{R.~Wood}
	\affiliation{Department of Physics, University of Surrey, Guildford GU2 7XH, United Kingdom}
	\author{P.J.~Woods}
	\affiliation{University of Edinburgh, Edinburgh EH9 3JZ, United Kingdom}
	\author{A.~Yeremin}
	\affiliation{Flerov Laboratory, Joint Institute for Nuclear Research, 141980 Dubna, Russia}


\date{\today}

\begin{abstract}
\noindent
\textbf{Background:} Previous measurements of $\beta$-delayed neutron emitters comprise around 230 nuclei, spanning from the $^{8}$He up to $^{150}$La. Apart from $^{210}$Tl, with a minuscule branching ratio of 0.007\%, no other neutron emitter is measured yet beyond $A=150$. Therefore new data are needed, particularly in the heavy mass region around N=126, in order to guide theoretical models and to understand the formation of the third r-process peak at $A\sim195$.\\
\textbf{Purpose:} To measure both, $\beta$-decay half-lives and neutron branching ratios of several neutron-rich Au, Hg, Tl, Pb and Bi isotopes beyond $N=126$.\\
\textbf{Method:} Ions of interest are produced by fragmentation of a $^{238}$U beam, selected and identified via the GSI-FRS fragment separator. A stack of segmented silicon detectors (SIMBA) is used to measure ion-implants and $\beta$-decays. An array of 30 $^3$He tubes embedded in a polyethylene matrix (BELEN) is used to detect neutrons with high efficiency and selectivity. A self-triggered digital system is employed to acquire data and to enable time-correlations. The latter are analyzed with an analytical model and results for the half-lives and neutron-branching ratios are derived using the binned Maximum-Likelihood method.\\
\textbf{Results:} Twenty new $\beta$-decay half-lives are reported for $^{204-206}$Au, $^{208-211}$Hg,$^{211-216}$Tl,$^{215-218}$Pb and $^{218-220}$Bi, nine of them for the first time. Neutron emission probabilities are reported for $^{210,211}$Hg and $^{211-216}$Tl.\\
\textbf{Conclusions:} The new $\beta$-decay half-lives are in good agreement with previous measurements in this region. The measured neutron emission probabilities are comparable or smaller than values predicted by global models like RHB+RQRPA.\\
\end{abstract}


\maketitle


\section{Introduction}
Very neutron-rich nuclei may emit one or more neutrons when they disintegrate via $\beta$ decay. This is the so-called $\beta$-delayed neutron ($\beta n$) emission process, which is energetically allowed when the Q$_{\beta}$-value of the decay exceeds the neutron separation energy (S$_{n}$) of the daughter nucleus. The $\beta n$-emission has been experimentally determined for about 230~neutron rich nuclei, spanning from $^{8}$He up to $^{150}$La~\cite{rudstam1993delayed,pfeiffer2002status}. Most of these measurements took advantage of the large fission yields around the two fission peaks at $A\sim95$ and $A\sim138$. However, $\beta$-delayed neutron emission has remained essentially inaccessible for nuclei heavier than $A=150$, where only a minuscule value of 0.007\% has been reported for the $\beta n$-emission probability of $^{210}$Tl~\cite{kogan1957neutron, stetter1962investigation}. Because of the scarce or non-existent $\beta n$ data in the heavy mass region, rapid neutron capture $r$-process~\cite{burbidge1957b2fh} calculations have to rely entirely on theoretical models~\cite{arcones2011dynamical,mumpower2016impact,surman2014sensitivity}. However, the performance of such models for reproducing the features of the $\beta$-decay in $r$-process waiting-point nuclei has been tested with experimental data only in the two shell-closures at $N=50$ and $N=82$ in measurements such as reported in ~\cite{ohm1980beta,kratz1981observation,kratz1982beta,gabelmann1982pnvalues,wang1999beta,montes2006beta,pereira2009beta}. The comparison is much more limited in the $N=126$ region, where only half-lives in the neighborhood of the doubly magic $^{208}$Pb were available (see e.g. Ref.~\cite{morales2014half}).

The neutron-rich nuclei ``south" of $^{208}$Pb are difficult to measure experimentally because of the very small production cross-sections and the large background conditions induced by the heavy primary beam. In the present work it was possible to produce and identify reliably secondary neutron-rich nuclei in the region ``south-east" of $^{208}$Pb in the chart of nuclides thanks to the high-energy (1~GeV/u) $^{238}$U beams available at the GSI facility. Extended motivation and results for this experiment were recently published in~\cite{Caballero-Folch2016First}. Here we present more details on the experimental apparatus in Section~\ref{sec:setup}, the analysis methodology and results, which are reported in Section~\ref{sec:analysis} and finally Sections~\ref{sec:discussion} and ~\ref{sec:conclusions} summarize the main results and conclusions.

\section{Experimental setup and ion identification}\label{sec:setup}
\begin{figure}[!ht]
\includegraphics[width =1.0 \columnwidth]{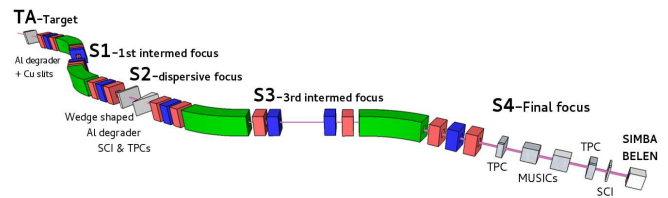}
\caption{(color online) Scheme of the FRS facility (see text for details).}
\label{fig:FRSscheme}
\end{figure}

The present measurements were carried out at the GSI Helmholtz Center for Heavy Ion Research. A $^{238}$U beam was accelerated to an energy of 1~GeV/u by the UNILAC linear accelerator coupled to the SIS 18 synchrotron. The average beam intensity was $2 \times 10^{9}$~ions/spill, with a pulsed beam structure of 1~s spill duration (SIS extraction) and a repetition cycle of 4~s. The beam impinged onto a $^9$Be target with a thickness of 1.6~g/cm$^2$ at the entrance of the Fragment Separator (FRS)~\cite{geissel1992gsi}. The selection of the ions of interest, from this point to the detection system, was done with the $B\rho-\Delta E - B\rho$-method using the FRS as an achromatic spectrometer. Neutron-rich nuclei ``south-east" of $^{208}$Pb in the chart of nuclides were produced using two FRS settings centered on $^{211}$Hg and $^{215}$Tl. In order to minimize the number of ionic charge states of the secondary beam a Niobium layer with a thickness of 223~mg/cm$^2$ was placed behind the Be-target. In addition, a homogeneous Al degrader with a thickness of 2.5~g/cm$^2$ was placed at the first FRS focal plane (S1) in combination with thick Cu slits, which served to reduce the contribution of fission fragments and primary beam charge-states with initially similar magnetic rigidity ($B\rho$) as the setting isotope. A wedge-shaped Al degrader with a thickness of 3874~mg/cm$^2$ was employed as achromatic degrader at the intermediate focal plane (S2), (see Fig.~\ref{fig:FRSscheme}).

Ion species were identified by means of standard FRS tracking detectors. Two plastic scintillators located at S2 and at the final focal plane (S4) were used to measure the time-of-flight ($t_\textsc{tof}$) of the ions. Two systems of Time Projection Chambers (TPCs)~\cite{janik2011time} placed at S4 and S2 allowed us to determine accurate $B\rho$ values for each ion by measuring their trajectory with respect to the central fragment. The measured $t_\textsc{tof}$ in combination with the $B\rho$ provided the necessary information to calculate the mass-to-charge ratio (A/q) on an event-by-event basis. The resolution thus obtained in A/q was 2.5$\permil$~\textsc{fwhm}.

In order to determine the atomic number ($Z$), two fast MUltiple Sampling Ionization Chambers (MUSICs)~\cite{MUSIC} were placed in the S4 experimental area. Although the detected nuclei were mainly bare, H- and He-like charge states were also detected in MUSICs. The latter events were corrected by combining the information of the two MUSICs and calculating the energy loss in the S2 degrader following the method applied in previous studies in this mass region~\cite{PhDAnabel,PhDCasarejos,PhDFabio,kurcewicz2012discovery}. In addition, it was needed to treat the gain fluctuations in the MUSIC detectors caused by the variations of the temperature in the experimental hall, and consequent changes in gas pressure, which were also corrected numerically~\cite{caballero2015tesi}. The final resolution obtained in Z for the Pb-Bi region was of $\lesssim 6\permil$~\textsc{fwhm}. Finally, the Z versus A/q particle identification diagram (PID) was experimentally validated with a dedicated run. $^{205}$Bi ions were implanted into a passive plastic stopper and the decay of well known isomeric transitions~\cite{kondev2004nuclear} were measured with HPGe detectors.

\begin{figure}[!ht]
\includegraphics[width = \columnwidth]{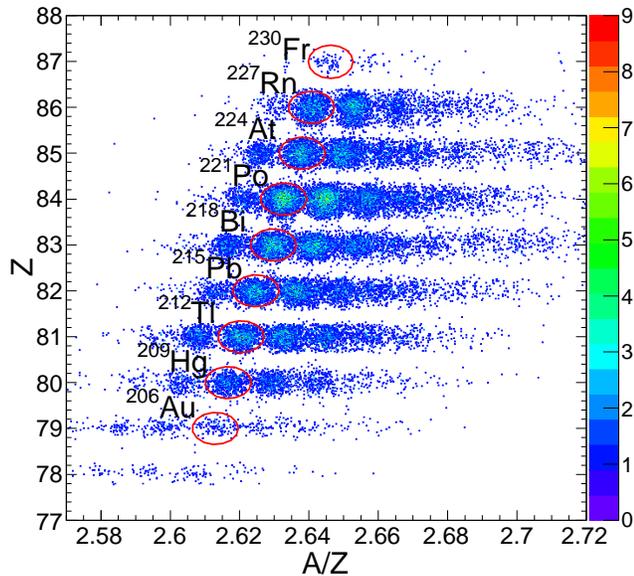}
\caption{(color online) Particle Identification Diagram (PID) with the total statistics of the $^{211}$Hg and $^{215}$Tl settings.}
\label{fig:PID}
\end{figure}

The PID obtained, including nuclei produced in both FRS settings with all the statistics accumulated during the experiment, is shown in Fig.~\ref{fig:PID}. All identified nuclei were already reported from previous experiments~\cite{alvarez2010production,chen2010discovery}. 

\subsection{Implantation and decay detection system}
\begin{figure}[!ht]
\centering
\includegraphics[width = 0.25\textwidth]{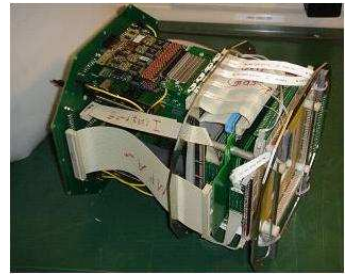}
\caption{(color online) Picture of SIMBA without its cylindrical coverage of 11.5~cm diameter.}
\label{fig:SIMBA2}
\end{figure}

\begin{figure}[!ht]
\centering
\includegraphics[width = 0.35\textwidth]{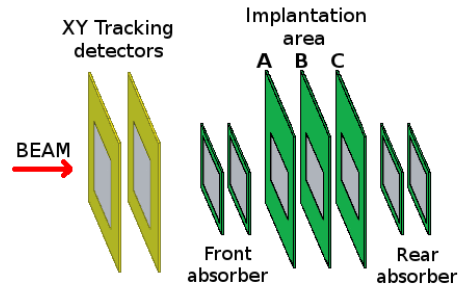}
\caption{Schematic view of SIMBA: From left to right (beam direction) the two $XY$-tracking silicons, the front absorber, the implantation layers $A$, $B$ and $C$ and the rear absorber layers. Adapted from~\cite{PhDSteiger,PhDHinke}.}
\label{fig:SIMBA}
\end{figure}

The secondary beam of neutron-rich isotopes was focused at the final focal plane (S4) of the FRS. A third Al degrader with variable thickness was adjusted in order to slow down the ions of interest and to implant them into the central region of an active stopper named ``Silicon IMplantation Beta Absorber" (SIMBA)~\cite{hinke2012superallowed, PhDSteiger}. SIMBA enabled position and energy measurement of heavy charged ions as well as charged particles from $\alpha$ and $\beta$ decays. The main difference between the SIMBA system employed here with respect to previous versions was the smaller number of Si detectors required to stop and implant the ions of the present experiment, which had a higher atomic number ($Z\sim82$) than those measured in the past ($Z\sim50$,~\cite{hinke2012superallowed}). In addition, the geometry and overall size of SIMBA were also modified in order to optimize neutron detection (see below). The present SIMBA version consisted of a stack of nine highly segmented Si detectors (see Figs.~\ref{fig:SIMBA2} and~\ref{fig:SIMBA}). 

\begin{figure}[!ht]
\centering
\includegraphics[width = \columnwidth]{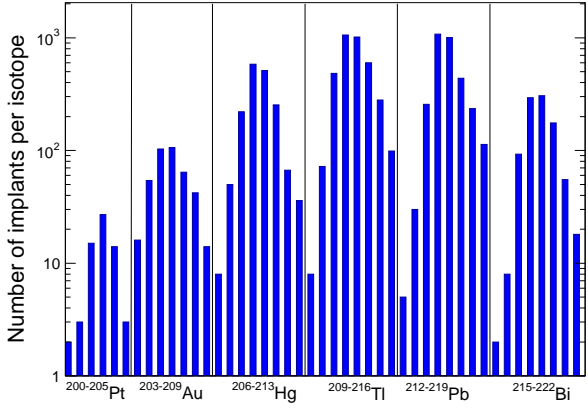}
\caption{Number of implanted nuclei of each isotope in the DSSSDs of SIMBA.}
\label{F3-NumImplants}
\end{figure}

The first two layers, called XY-tracking detectors, consisted of single-sided Si-strip detectors (SSSD) with their strips in orthogonal orientation with respect to each other. They were used for determining accurately the ion position in the transversal plane of the beam. A center-of-gravity method applied to the charge shared over all the strips allowed us to determine the ion position with an accuracy of $\pm 1 mm^2$~\cite{PhDKarl,caballero2015tesi}, corresponding to one pixel in a silicon layer of SIMBA. The implant and decay sensitive region consisted of two SSSD layers (front absorbers), three double sided silicon stripped detectors (DSSSD) designated as implantation layers A, B and C, and two SSSD layers (rear absorbers). The energy deposited by the ions along these seven Si detectors was used to detect whether the ion was implanted or if it punched through, as well as to determine the corresponding implant layer or depth. The segmentation of the DSSSDs was 60-fold in X and 40-fold in Y direction, with a strip width of 1~mm. Fig.~\ref{F3-NumImplants} shows the total amount of implanted ions for each isotope in the DSSSDs of SIMBA.

\begin{figure}[!ht]
\centering
\includegraphics[width =0.49 \columnwidth]{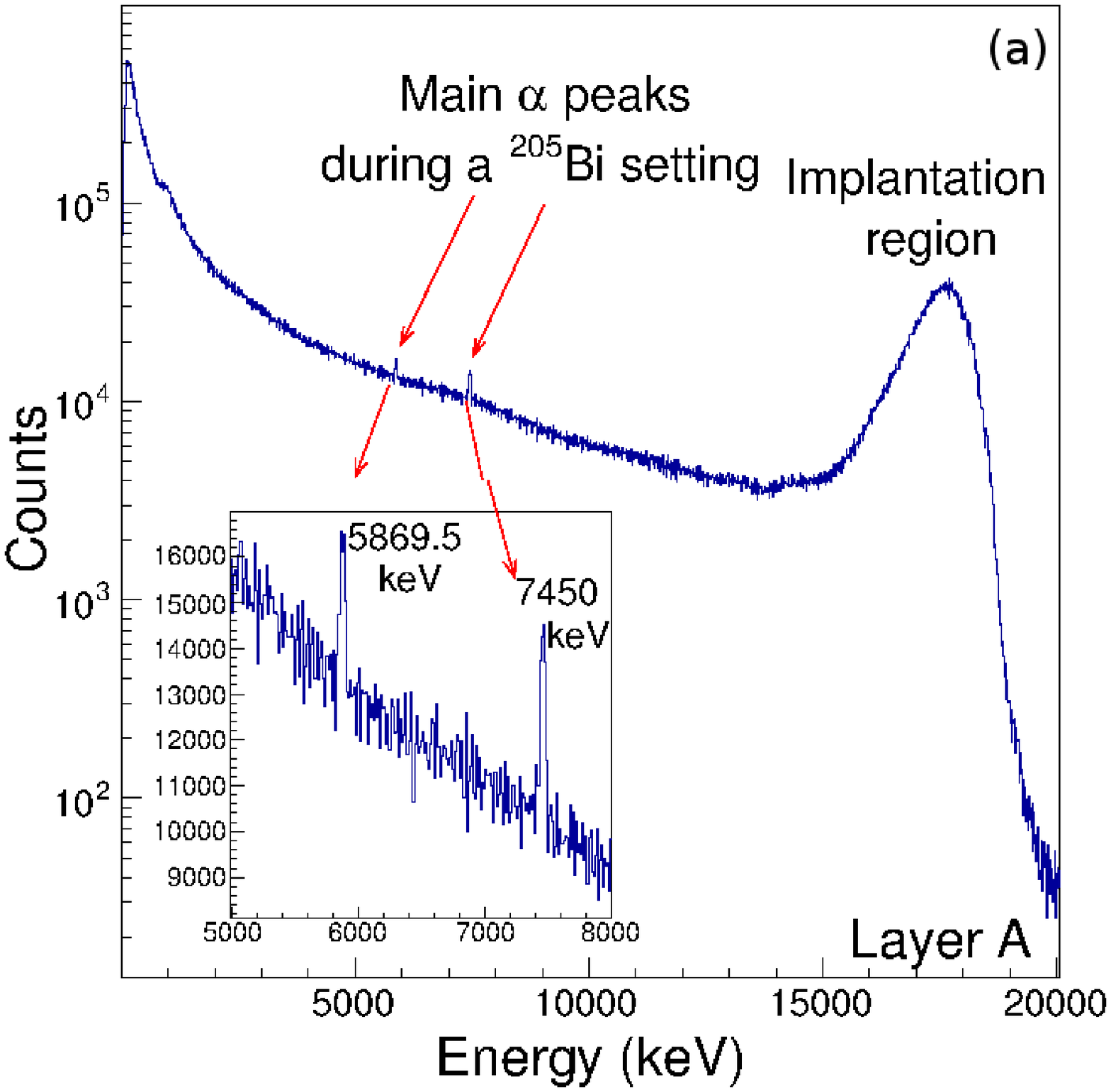}
\hfill
\includegraphics[width =0.49 \columnwidth]{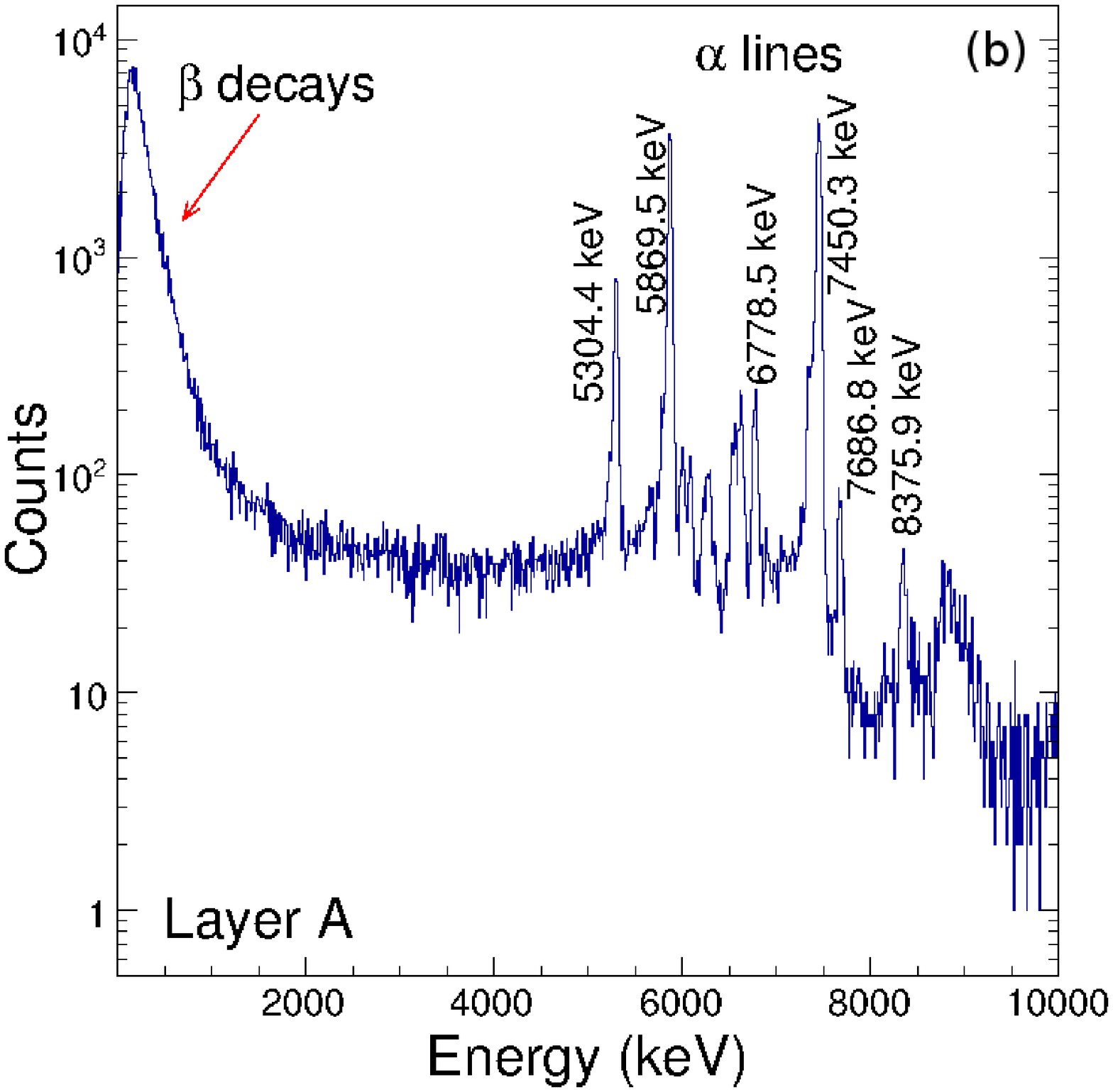}
\includegraphics[width =0.49 \columnwidth]{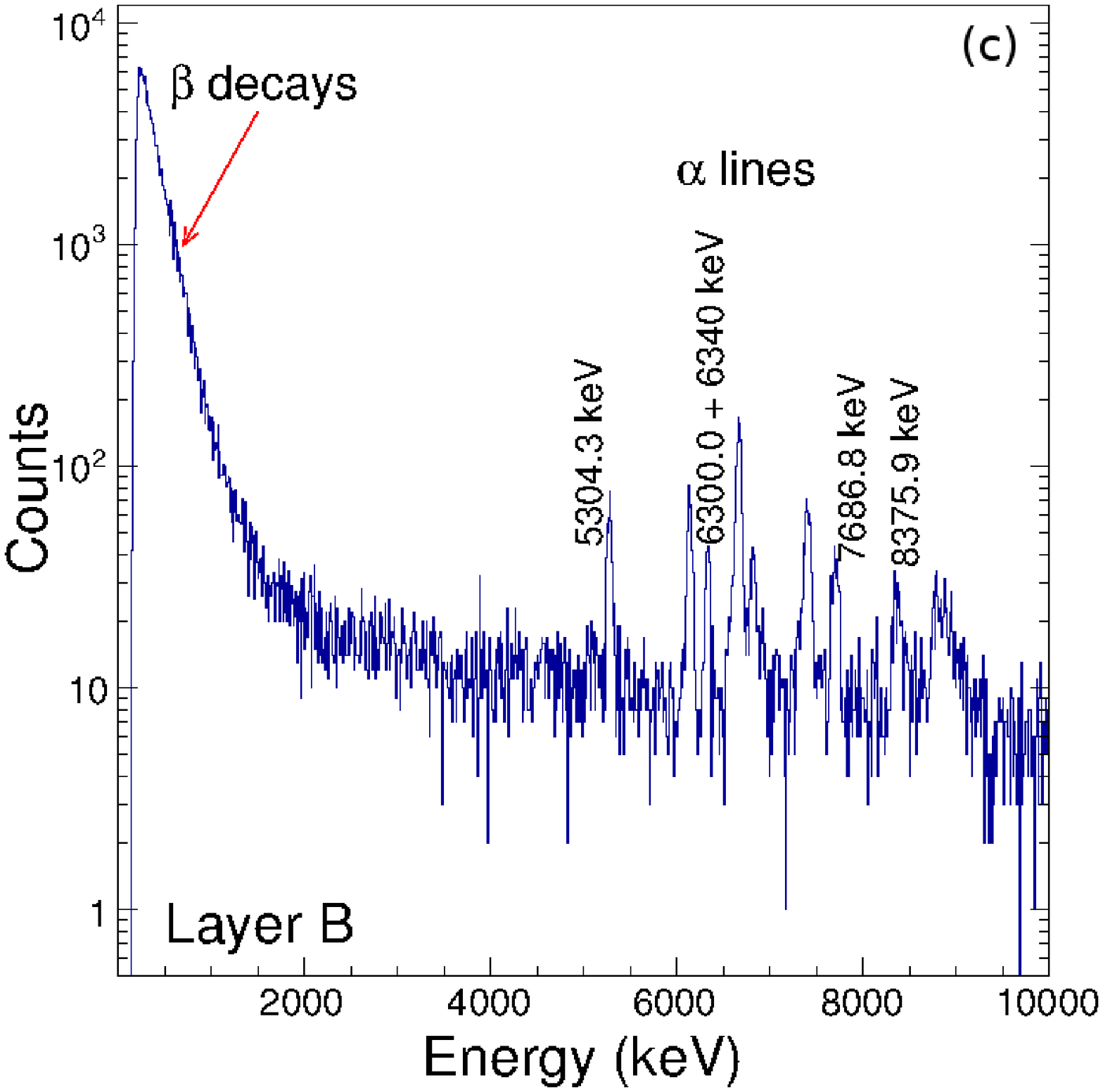}
\hfill
\includegraphics[width =0.49 \columnwidth]{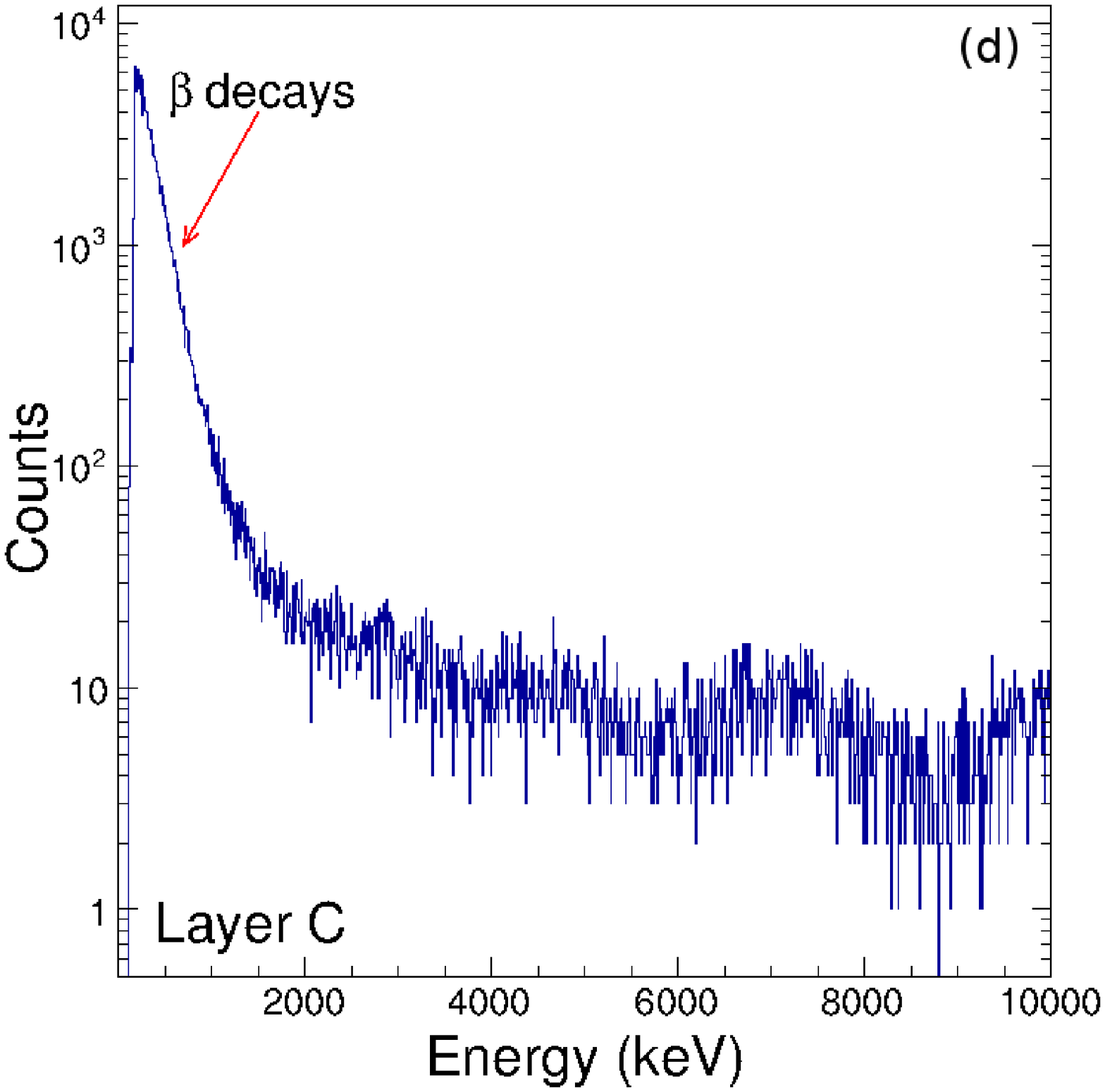}
\caption{Calibrated energy spectra for SIMBA layers: A (top), B and C (bottom).}
\label{fig:dssd}
\end{figure}

The energy deposited by $\beta$ decays in each Si-electrode showed, as expected, a continuum spectrum which extended up to $\sim$2-3~MeV. The energy deposited in the DSSSDs was determined from the response of the Y-strips, which were readout via logarithmic preamplifiers. The latter allowed to clearly separate the ion implants and decays in the spectrum (see Table~\ref{T2-Alpha} and Figs.~\ref{fig:dssd}a and~\ref{fig:dssd}b). A pulse generator was used to perform the gain-matching of the Y-strips. An accurate energy calibration of layers A and B was accomplished using well known $\alpha$ decays from several At, Bi, Rn and Po isotopes produced in the decay of implanted Tl, Pb and Bi nuclei (see Figs.~\ref{fig:dssd}a and ~\ref{fig:dssd}b). No $\alpha$ emitter was implanted in the layer C and, therefore, a coarse energy calibration of the latter was made using only the broad $\beta$ spectrum (Fig.~\ref{fig:dssd}d). Nevertheless, for implant-$\beta$ time-correlations only an energy window covering the broad $\beta$ spectrum is required and, therefore, an accurate energy calibration is of secondary relevance for this data analysis.

\begin{table}[!ht]
  \caption{$\alpha$ lines observed in SIMBA in layers A and B and their associated nuclei (see some of them in Figs.~\ref{fig:dssd}b and ~\ref{fig:dssd}c).\label{T2-Alpha}}
  \begin{tabular}{ccccc}
\hline
\\
$\alpha$ energy &SIMBA &$\alpha$ emitter&Precursor\\
(keV) &layer &&implanted\\
\hline
\\
5304.3 &A, B &$^{210}$Po& (Implanted)\\
5869.5 &A&$^{211}$At& (Implanted)\\
6002.4 &A&$^{218}$Po&$^{218}$Bi, $^{218}$Pb\\
6050.8 &A&$^{212}$Bi&$^{212}$Tl\\
6208.0 &A&$^{219}$At&$^{219}$Bi\\
6288.1 &A&$^{220}$Rn&$^{220}$Bi\\
6300.0 &B&$^{212}$Bi&$^{212}$Tl\\
6340.0 &B&$^{212}$Bi&$^{212}$Tl\\
6537.0 &A&$^{217}$Po&$^{217}$Pb\\
6622.9 &A&$^{211}$Bi&$^{211}$Tl\\
6778.5 &A&$^{216}$Po&$^{216}$Pb\\
7386.1 &A&$^{215}$Po&$^{215}$Pb\\
7450.3 &A&$^{211}$Po&$^{211}$At\\
7686.8 &A,B&$^{214}$Po&$^{214}$Pb, $^{214}$Tl\\
8375.9 &A,B&$^{213}$Po&$^{213}$Tl\\
\hline
\hline
\end{tabular}
\end{table}

\begin{figure}[!htbp]
\centering
\includegraphics[width = 0.51\columnwidth]{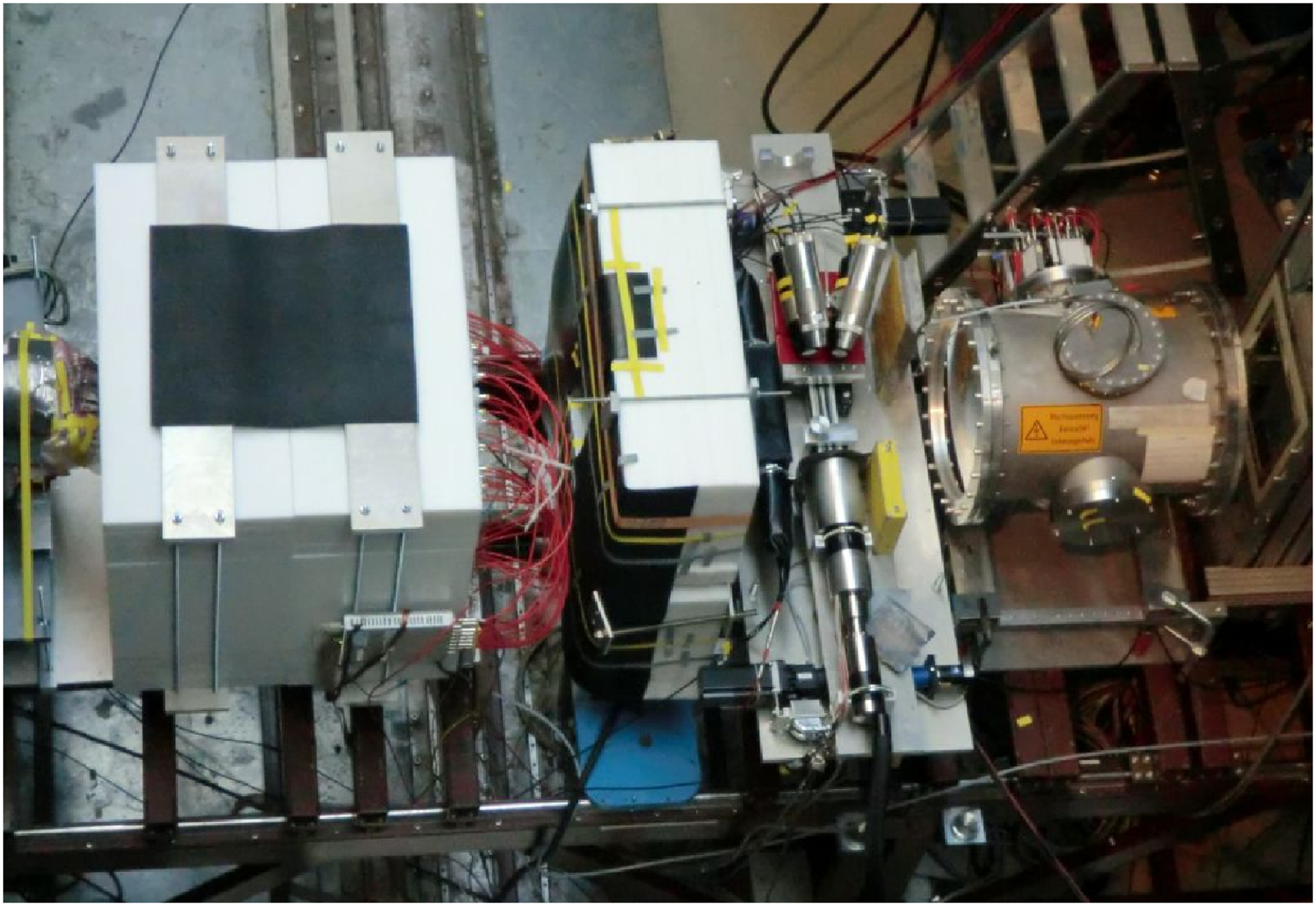}
\hfill
\includegraphics[width = 0.4675\columnwidth]{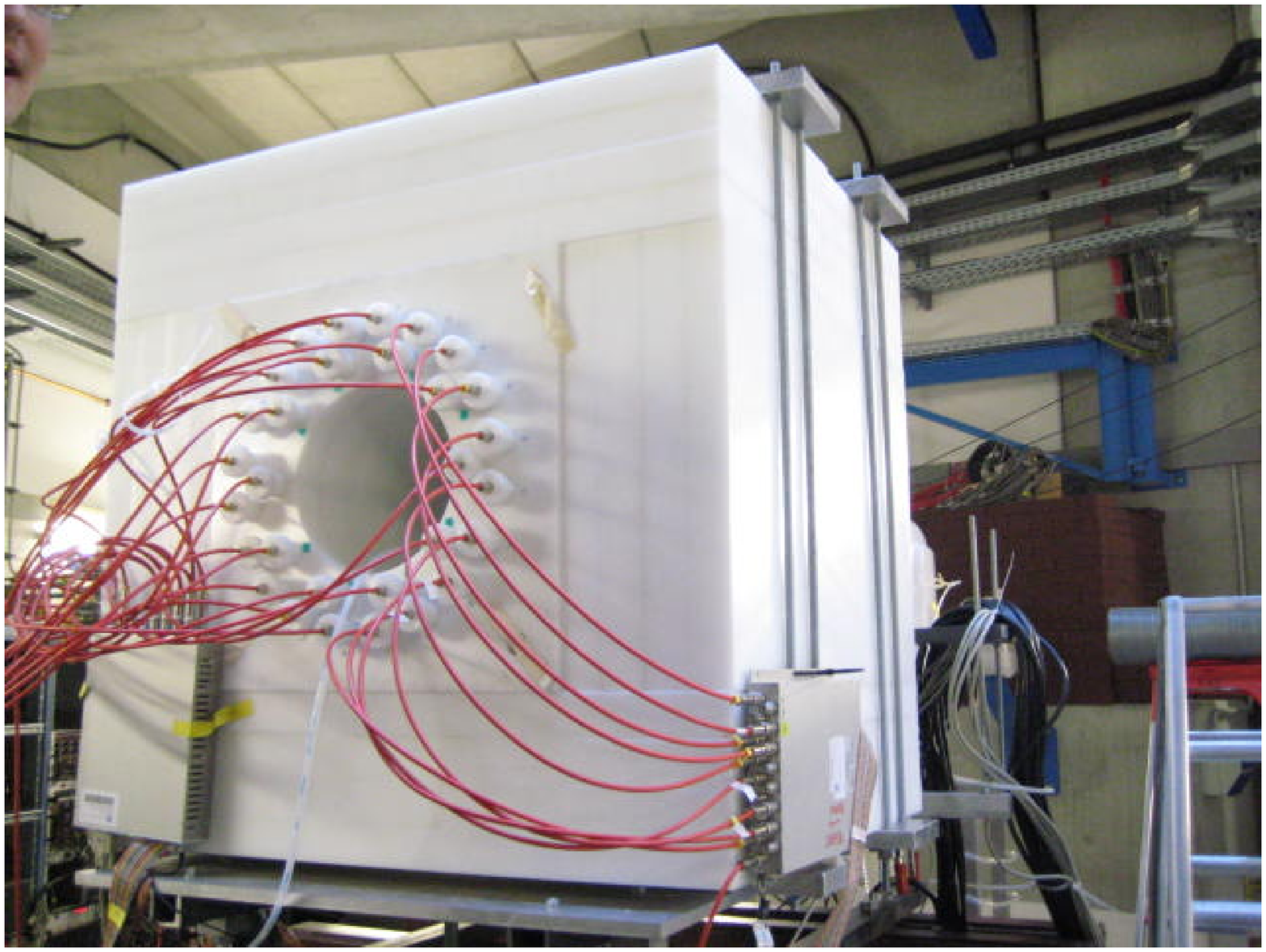}
\caption{Left: Top view of BELEN and the front PE-shielding wall (beam enters from the right-hand side). Right: Picture of the BELEN detector showing the two rings of counters, the central hole where SIMBA (see Fig.~\ref{fig:SIMBA2}) was placed and the extra 20~cm of PE shielding around.}
\label{fig:setup_picture}
\end{figure}

\begin{figure}[!ht]
\centering
\includegraphics[width = 0.5\textwidth]{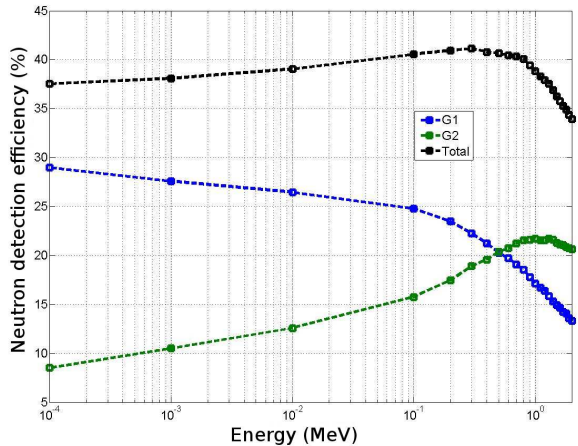}
\caption{MCNPX simulation of the neutron detection efficiency of BELEN as a function of the neutron energy. G1 and G2 refer to the two rings of $^3$He counters. See text for details.}
\label{fig:efficiency}
\end{figure}

SIMBA was placed inside the cylindrical hole (23~cm diameter) of the Beta dELayEd Neutron (BELEN) detector~\cite{BELEN-hyperfine,agramunt2014new,PhDGorlychev,TDRBELEN}. BELEN consisted of an array of 30 $^3$He-counters of 2.54~cm diameter, embedded in a high-density polyethylene (PE) matrix (Fig.~\ref{fig:setup_picture}).
The $^3$He tubes were distributed in two rings, an inner one with a radius of 14.5~cm and 10~tubes of 10~atm, and the outer ring with a radius of 18.5~cm and 20 tubes of 20~atm. This configuration was designed by means of \textsc{Geant4}~\cite{PhDRiego,agostinelli2003geant4} and MCNPX~\cite{pelowitz2005mcnpx,fishman1996algorithms, rubinstein1981simulation} Monte Carlo (MC) simulations in order to achieve a high and flat detection efficiency (see Fig.~\ref{fig:efficiency}). Up to a neutron energy of 1~MeV the detection efficiency was 40(2)\%, and it decreased to 25\% at 5~MeV. The MC codes were experimentally validated at E$_{n}$=2.3~MeV with a dedicated measurement of a well calibrated $^{252}$Cf source.

\begin{figure}[!htbp]
\centering
\includegraphics[width = 0.8\columnwidth]{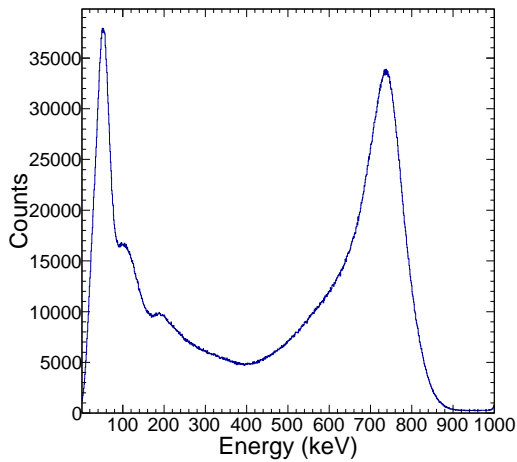}
\caption{Energy spectrum measured with all 30 $^3$He tubes of BELEN during the $^{211}$Hg setting.}
\label{fig:response}
\end{figure}

The Q$_{\beta n}$ window of the exotic nuclei involved in the present measurement spans neutron energies up to 2.5~MeV (see Table~\ref{T5-Pn}). In this energy range, the average neutron detection efficiency is 38\%; this value was used in the data analysis (see Sec.~\ref{sec:analysis}). Using the calibrated $^{252}$Cf source a gain-matching of the response of all 30~counters was carried out before the experiment. The stability of the overall detector response was checked regularly during the experiment. The accumulated spectrum for all 30 tubes during the $^{211}$Hg setting is shown Fig.~\ref{fig:response}.

The energy window considered in the data analysis as neutron events comprises the range from a low threshold of the 191~keV peak up to the end of the main peak of the reaction (see Eq.~\ref{E3_nReac}) at 765~keV. This peak corresponds to the kinetic energy realized by the two reaction products, a triton and a proton.
\begin{equation}
\centering
^{3}He + n \longrightarrow~^{3}H +~^{1}H + 765~keV
\label{E3_nReac}
\end{equation}

In order to reduce the neutron background in BELEN two additional elements were implemented in the setup. On one hand, a PE wall with a thickness of 30~cm was installed upstream from BELEN in order to shield the detector from neutron background induced by the secondary beam (see Fig.~\ref{fig:setup_picture}). This wall had a central hole of 180~mm in $X$-direction and 70~mm in $Y$-direction to let the beam into SIMBA. In addition, a layer of borated rubber was attached to the back side of the PE wall in order to absorb thermalized neutrons that could eventually reach the rear side of the wall. On the other hand, the BELEN detector was surrounded by 20~cm of PE shielding (see Fig.~\ref{fig:setup_picture}) in order to moderate and absorb scattered neutrons from the surroundings. 

The GSI Multi Branch System (MBS)~\cite{MBSweb} was used to acquire data from the FRS tracking detectors and SIMBA. This data acquisition system (DACQ) was triggered by a scintillator at S4 with an efficiency of $\sim$100\% for heavy ions. 
MBS was also triggered by high-energy implant and low-energy $\beta$-decay events in SIMBA. On the other hand, the neutron data from BELEN were acquired using a digital self-triggered system~\cite{agramunt2016characterization} based on SIS3302 VME digitizers from Struck Innovative Systems~\cite{Struck-SIS3302web}. 
Each module had 8 input channels running at 100~MHz sampling rate with an ADC resolution of 16~bit. 
A common clock was used for time-stamping the events acquired in the BELEN and MBS DACQ systems with 20~ns resolution. In this way, ion-implant versus $\beta$-decay time-correlations and ion-$\beta$-neutron time-correlations could be built over an arbitrarily long time-window and in both, forward (increasing) and backward (decreasing) time directions. The latter aspect was a key feature in the analysis stage in order to determine reliably the background level (see Sec.~\ref{sec:analysis}).

\section{Determination of $\beta$-decay half-lives and P$_{1n}$ values}\label{sec:analysis}
The methodology followed here for the analysis of the $\beta$-decay half-lives is similar to the approach successfully applied in previous experiments at ISOL and fragmentation facilities, such as described in~\cite{montes2006beta,hinke2012superallowed}. There are two fundamental aspects in this data analysis, namely a reliable background characterization of $\beta$- and neutron events and the spatial- and temporal- correlation approach. We describe first the analysis methodology from a rather general perspective and afterwards we show in detail its application to one of the largest implant statistics case, which is $^{213}$Tl. The latter is also used to illustrate the background treatment in the analysis.

Let us consider one specific isotope $i$. The Bateman analytical expression~\cite{Bateman1910equations} describes the time-evolution of its abundance $N_i(t)$. In this analysis we assumed that only the parent ($N_1$) and daughter ($N_2$) decays are contributing to the decay curve. This assumption is justified because all granddaughters of the analyzed nuclei are either stable nuclei or have a half-life much longer than that of parent and daughter nuclei. The resulting expression after this assumption is given by
\begin{equation}
\sum\limits_{i=1}^{2}\lambda_iN_i(t) = (\lambda_1N_1(t) + \lambda_2N_2(t)),
\label{E4-Bateman}
\end{equation}
where $\lambda_1 = ln(2)/T_{1/2}$ is the decay constant for the implanted nucleus $(i=1)$, with unknown half-life $T_{1/2}$, and $\lambda_2$ the decay constant of the daughter nucleus $(i=2)$. As it is shown below, in many cases even the contribution of the daughter nucleus was very small.

Regarding the spatial correlation, we considered an implant and a decay event to be spatially associated when the $\beta$-decay position measured in layers A, B or C of SIMBA was within a correlation region of 3~mm$^2$ around the implant position measured in the same layer. Extending this condition to neighboring layers did not improve noticeably the statistics and was therefore disregarded. In general, smaller and larger correlation areas were not found to provide a better result in terms of statistics and signal-to-background ratio in the time-correlation diagrams~\cite{caballero2015tesi}.

Regarding the time correlation, the methodology followed here consisted of building a correlation histogram for every implant of a certain species $i$ containing its time difference with respect to all subsequent $\beta$ events within a broad time window $\Delta_t \gg T_{1/2}$. In this analysis we used ten times the expected half-life, $\Delta_t \simeq 10\cdot T_{1/2}$. As it is demonstrated below, the uncorrelated background rate is a constant value, that can be referred to as $b$. In this case, the probability density function describing the time-dependency of the correlation distribution is given by~\cite{bernas1990beta}
\begin{equation}
\begin{split}
\rho(\lambda_1,t) =& \varepsilon_{\beta} b + \varepsilon_{\beta} \lambda_1 \mathrm{e}^{-\lambda_1 t}\\
+& \varepsilon_{\beta} \frac{\lambda_1 \lambda_2}{\lambda_1 - \lambda_2} \big(\mathrm{e}^{-\lambda_2 t} - \mathrm{e}^{-\lambda_1 t}\big),
\label{E4-Allbetas}
\end{split}
\end{equation}
where $\varepsilon_{\beta}$ is the $\beta$-detection efficiency in SIMBA. Thus, for a certain number of implanted events $N_{1}(0)$, the total number of $\beta$ particles detected $N_{\beta}$ at a time $t$ with respect to the implantation time ($t=0$) is given by 
\begin{equation}
\begin{split}
N_{\beta}^{All \beta}(t) =& N_{1}(0) \cdot \rho(\lambda_1,t) \cdot \Delta t = \\
=& \varepsilon_{\beta} \cdot (\lambda_1N_1(t) + \lambda_2N_2(t) + b) \cdot \Delta t,
\label{E4-Allbetas-2}
\end{split}
\end{equation}
where $N^{All \beta}_{\beta}(t)$ is the total number of detected decays at a time $t$, $b$ is the $\beta$-background normalized and corrected by $\varepsilon_{\beta}$ and $\Delta t$ corresponds to the bin time-width used in the implant-$\beta$ time-correlation histogram. The time evolution of the parent abundance is described by $N_1(t) = N_{1}(0)\cdot\mathrm{e}^{-\lambda_1 t}$, whereas the contribution of the daughter is given by $N_2(t) = N_{1}(0)\cdot\frac{\lambda_1 }{\lambda_1 - \lambda_2} \big(\mathrm{e}^{-\lambda_2 t} - \mathrm{e}^{-\lambda_1 t}\big)$, assuming $N_{2}(0) = 0$. In the data analysis the common factor $N_{1}(0)\cdot\varepsilon_{\beta}$ in Eq.~\ref{E4-Allbetas-2} is derived from the first bin in the correlation diagram~\cite{caballero2015tesi}. In this way the quantity of interest ($T_{1/2}$) can be reliably determined without knowing explicitly the $\beta$-detection efficiency.

\subsubsection*{Reference analysis of the $^{213}$Tl half-life}
\begin{figure}[!htbp]
\includegraphics[width = 0.78 \columnwidth]{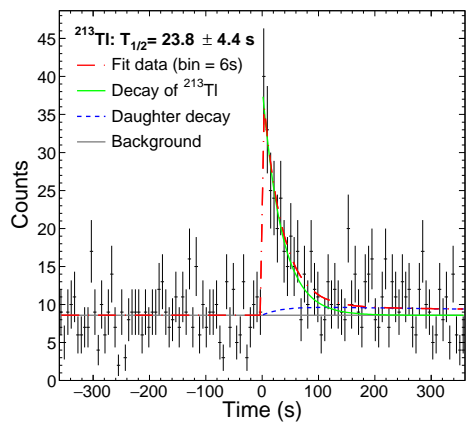}
  \caption{$^{213}$Tl implant-$\beta$ time-diagram using a correlation area of 3$\times$3~mm$^2$ and $\beta$-events detected outside of the spill. The red line shows the result of a binned Maximum Likelihood (ML) analysis. The decay curve shows mostly the contribution of $^{213}$Tl decay (green line), with a minor contribution from the daughter $^{213}$Pb (dashed blue line).}
\label{F4-213Tl}
\end{figure}

The thallium isotope $^{213}$Tl was implanted with large statistics (1015 implants) and therefore we used this case to establish the analysis methodology on a reliable statistical basis. The $\beta$ background showed a dependency with the time structure of the pulsed primary beam. Indeed, during beam extraction from SIS (1~s) the overall background level of $\beta$-like events in SIMBA was $\sim$40\% higher than during the time-interval between spills~\cite{caballero2015tesi}. This feature led to a better signal-to-background ratio in the implant-$\beta$ time-correlation histograms when only $\beta$-events outside of the spill time-intervals were considered in the correlations, when compared to the same diagram including $\beta$-events inside and outside of the spill. With this restriction in mind, the background level evaluation was based on time backward ($t<0$) implant-$\beta$ correlations, i.e. time difference between each implant and all the $\beta$-events occurred before it, within a broad time window ($\Delta t \simeq 10 \cdot T_{1/2}$) and in the same correlation area used in the forward analysis of 3$\times$3~mm$^2$. The background level thus determined allows one to adjust the parameter $b$ in Eq.~\ref{E4-Allbetas}. This approach is illustrated in Fig.~\ref{F4-213Tl} for the case of $^{213}$Tl, which shows backward (negative) and forward (positive) implant-$\beta$ correlations. The contribution to the measured $\beta$-activity from decays of other nuclei can be assumed to be negligible due to the very low average implantation rate of $2 \times 10^{-5}$~ions/s/pixel.
Using Eq.~\ref{E4-Allbetas-2} a binned Maximum Likelihood (ML)~\cite{proakis1994communication} analysis of the time correlation histogram was carried out, which yielded a half-life for $^{213}$Tl of $T_{1/2}=23.8\pm4.4$~s. Comparison with other literature values is detailed in Table~\ref{T5-HL}. 
\subsubsection*{Neutron branching ratio for $^{213}$Tl}
\begin{figure}[!ht]
  \includegraphics[width = 0.78 \columnwidth]{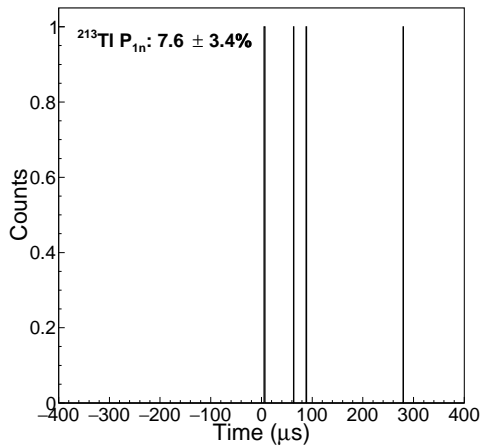}
  \caption{$\beta$-neutron correlation events during the $^{213}$Tl implant-$\beta$ correlation time.}
  \label{F4-neu-1}
\end{figure}

The implant-$\beta$-neutron correlations were analyzed by selecting a correlation window of $\Delta^n_t=$400~$\mu s$ forward and backward in time, following each $\beta$ detection. This time interval was determined according to the expected neutron moderation time in polyethylene. Considering the $\beta$ efficiency $\varepsilon_{\beta}$ as a constant value along the range of energies of interest, the $P_{1n}$-value can be directly obtained from the subtraction of time-forward and time-backward $\beta$-neutron correlated events, 
\begin{equation}
	P_{1n} (\%) = \frac{1}{\varepsilon_{n}} \frac{N_{\beta n}^{fwd} - N_{\beta n}^{bkd}}{N_{\beta}} \cdot 100
	\label{E4-Pn-equation}
\end{equation}
where $\varepsilon_{n}$ is the BELEN neutron efficiency, $N_{\beta n}^{fwd}$ the number of forward correlated implant-$\beta$-neutron events and the $N_{\beta}$ the number of parent $\beta$-decays. $N_{\beta n}^{bkd}$ designates the backward $\beta$-neutron correlations, which were used to define the uncorrelated neutron background level. The efficiency can be considered flat along the energy range of interest according to the $Q_{\beta n}$-values of the implanted isotopes (100~keV - 2.5~MeV, see Table~\ref{T5-Pn}), with a constant value of 38\% and with a relative uncertainty of 5\% (see Fig.~\ref{fig:efficiency}). This overall uncertainty includes the contributions of statistical errors of $N_{\beta n}^{fwd}$, $N_{\beta n}^{bkd}$ and $N_{\beta}$ together with the uncertainty for the BELEN detector efficiency from the fluctuations along the energy range. For $^{213}$Tl we measured 5 forward and none backward correlated neutrons (Fig.~\ref{F4-neu-1}), which yielded a $P_{1n}$-value of $7.6\pm3.4~\%$.

\subsection{Thallium isotopes: $^{211-216}$Tl}
\begin{figure*}[!htbp]
  \centering
  \includegraphics[width = 0.38 \textwidth]{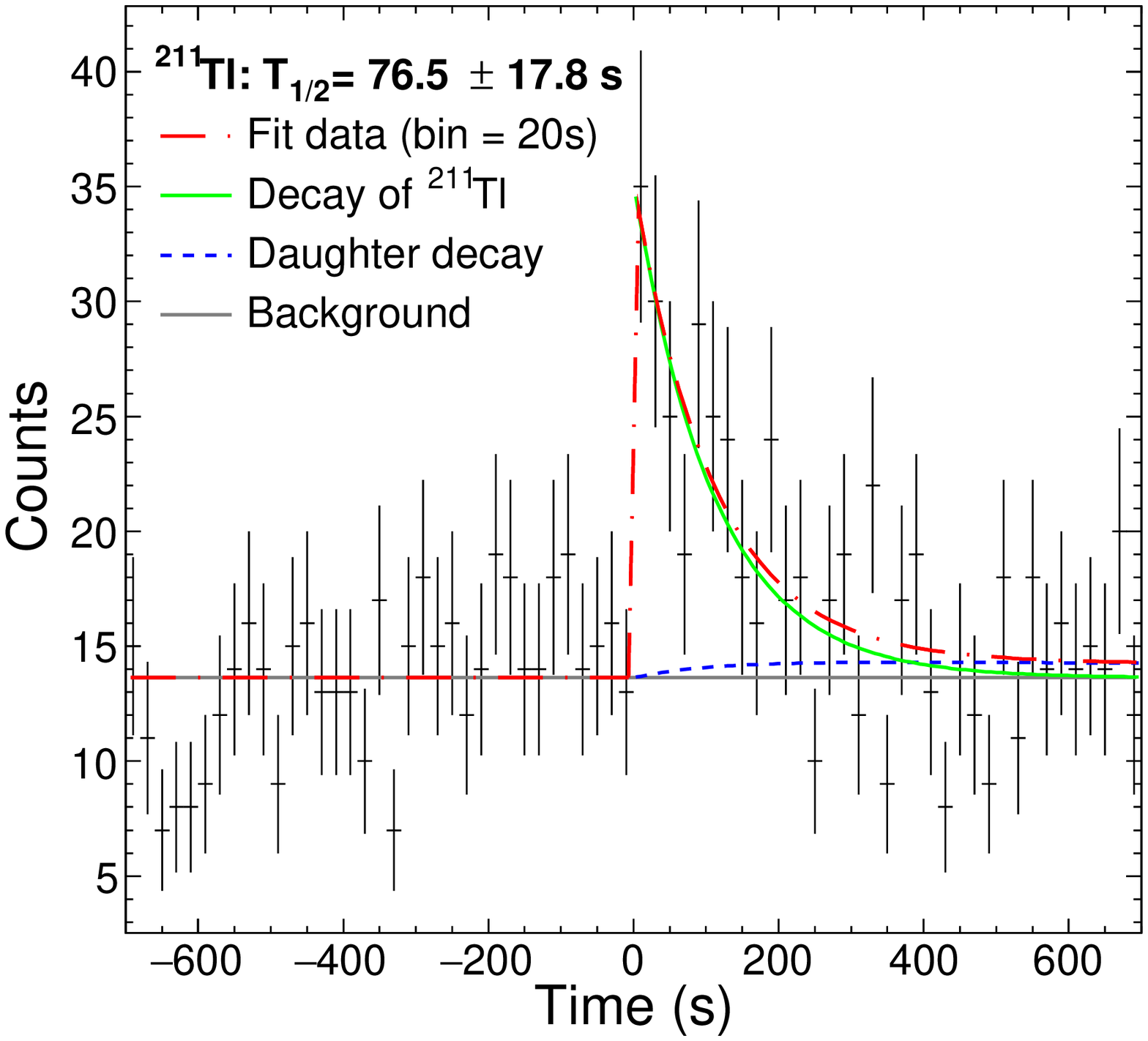}
  \includegraphics[width = 0.38 \textwidth]{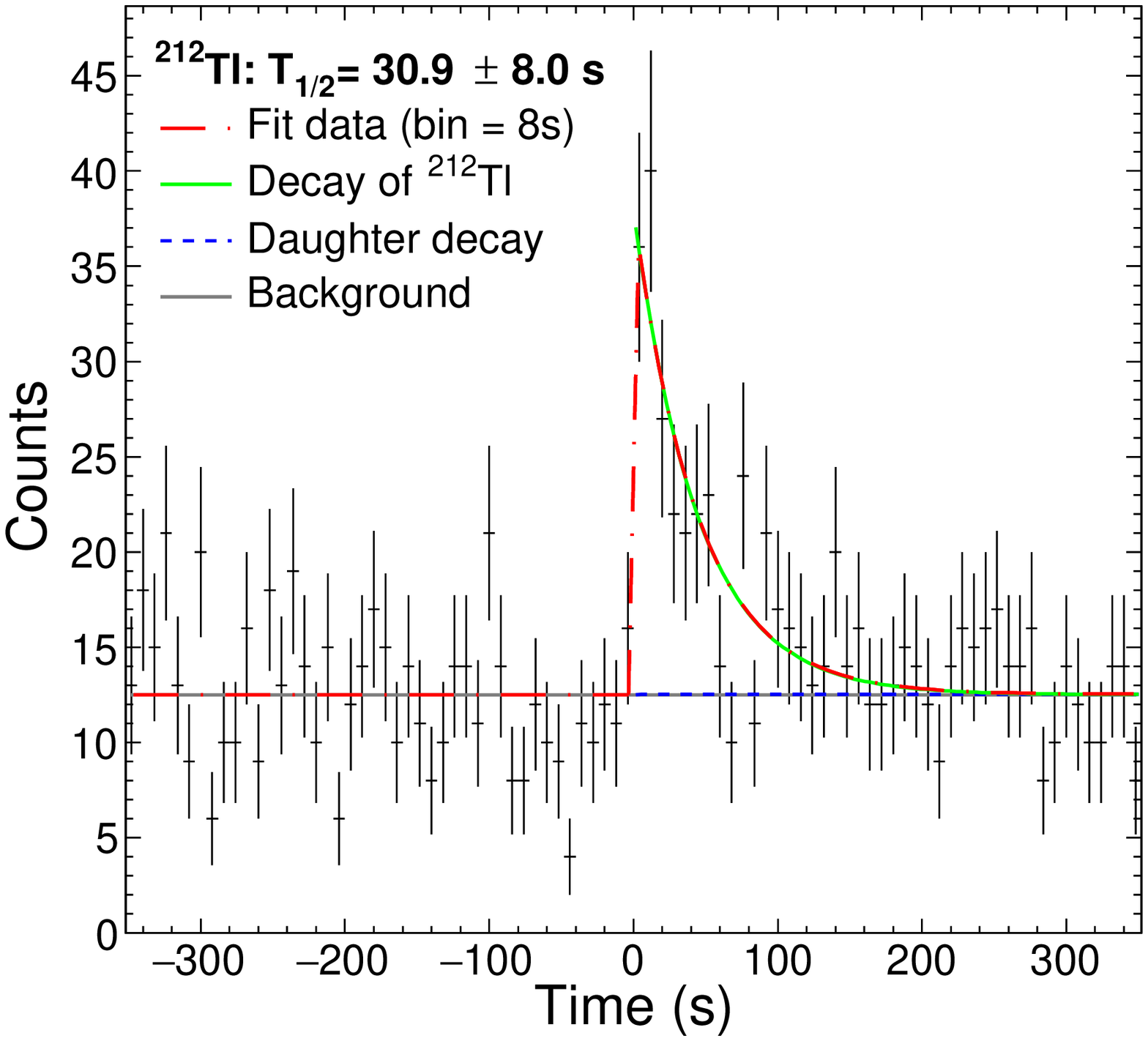}
  \includegraphics[width = 0.38 \textwidth]{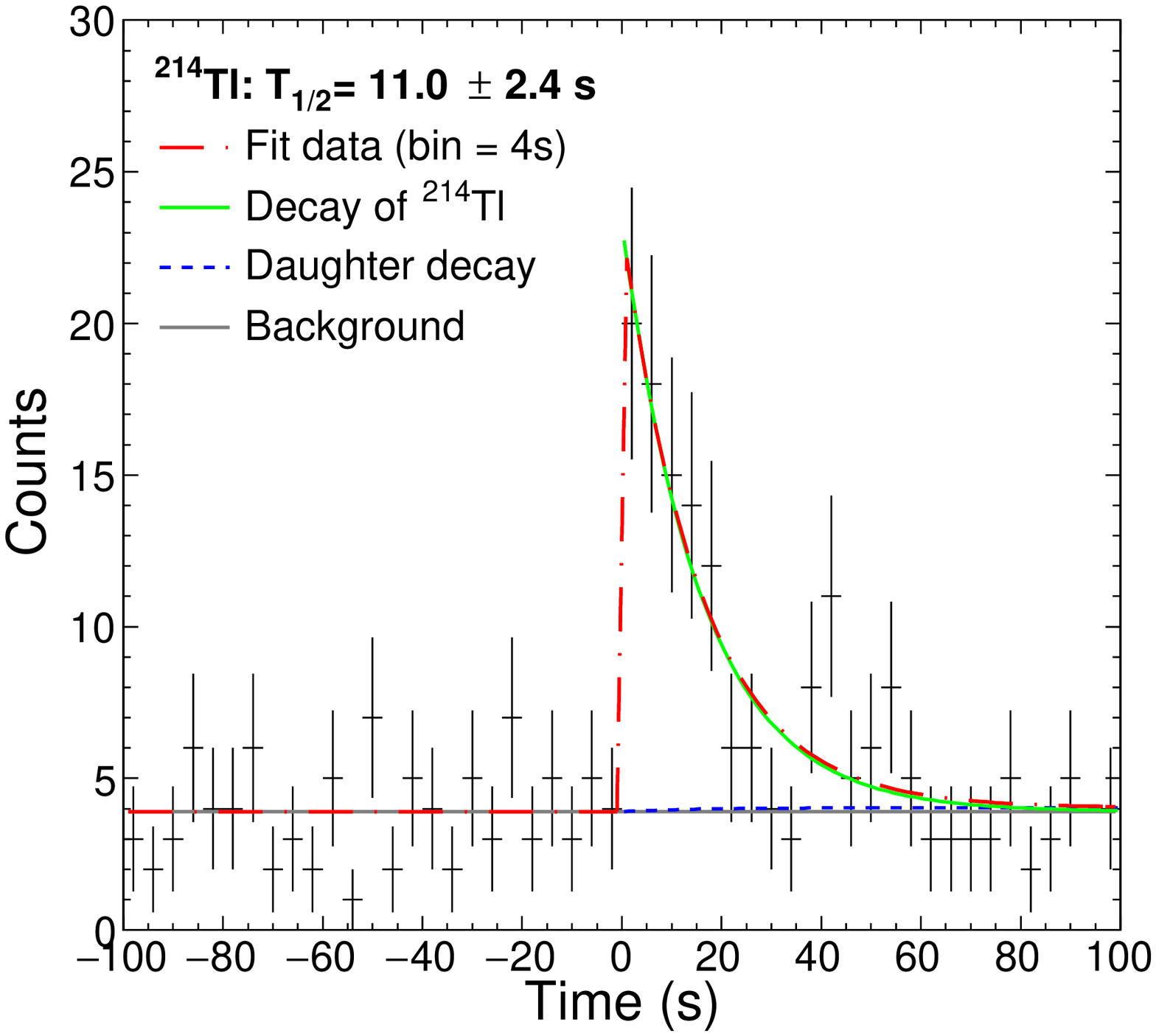}
  \includegraphics[width = 0.38 \textwidth]{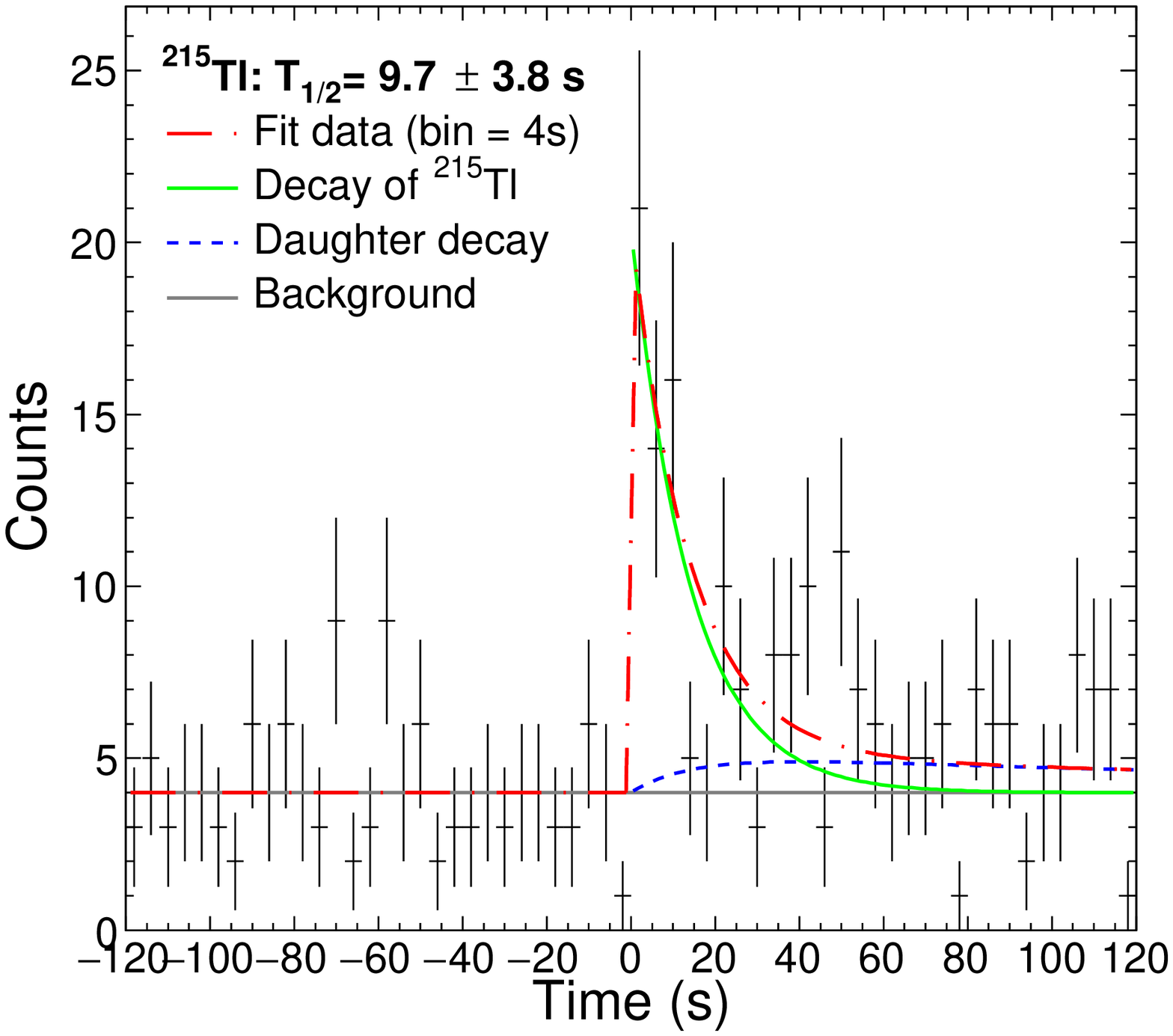}
  \includegraphics[width = 0.38 \textwidth]{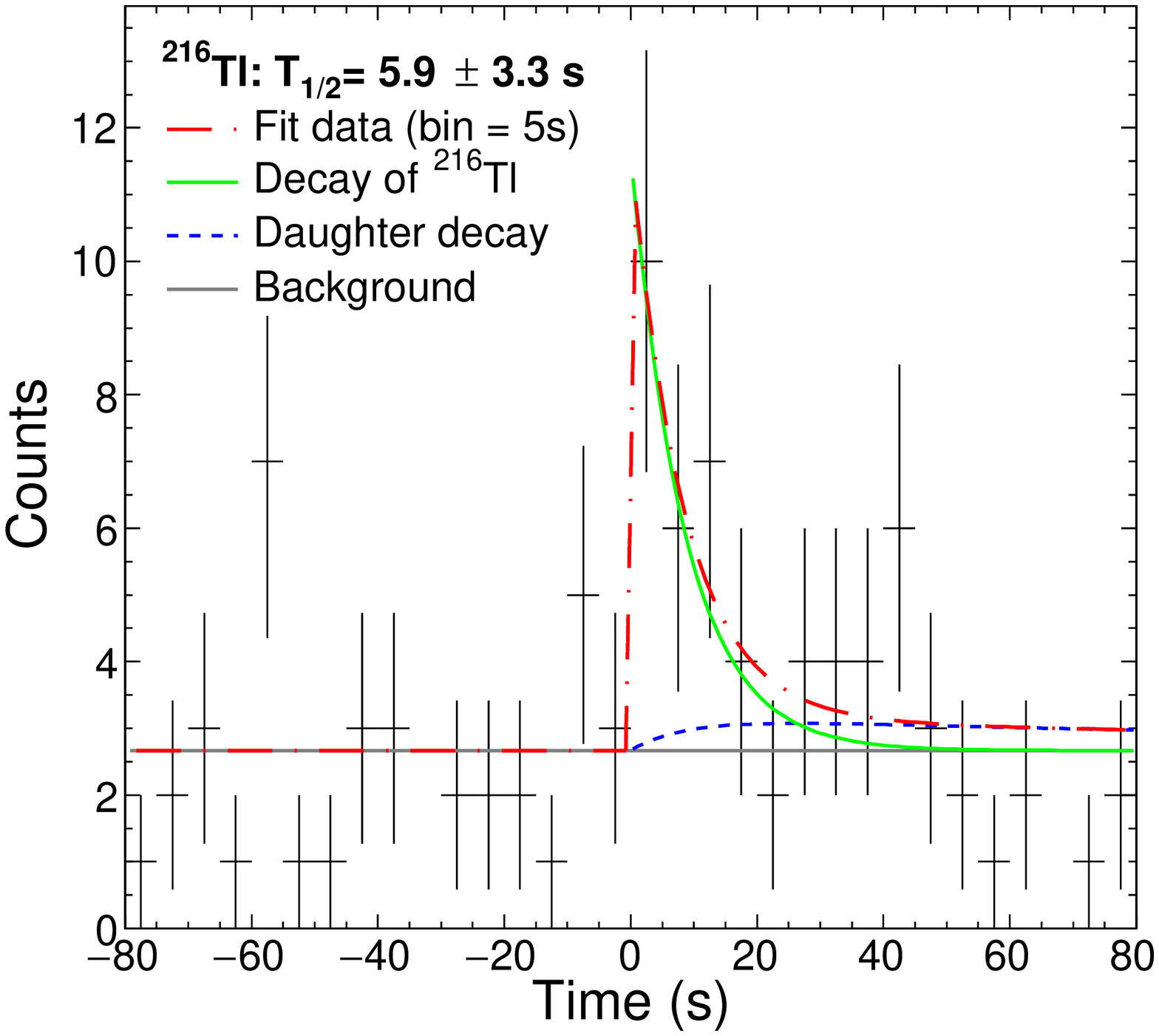}
  \includegraphics[width = 0.38 \textwidth]{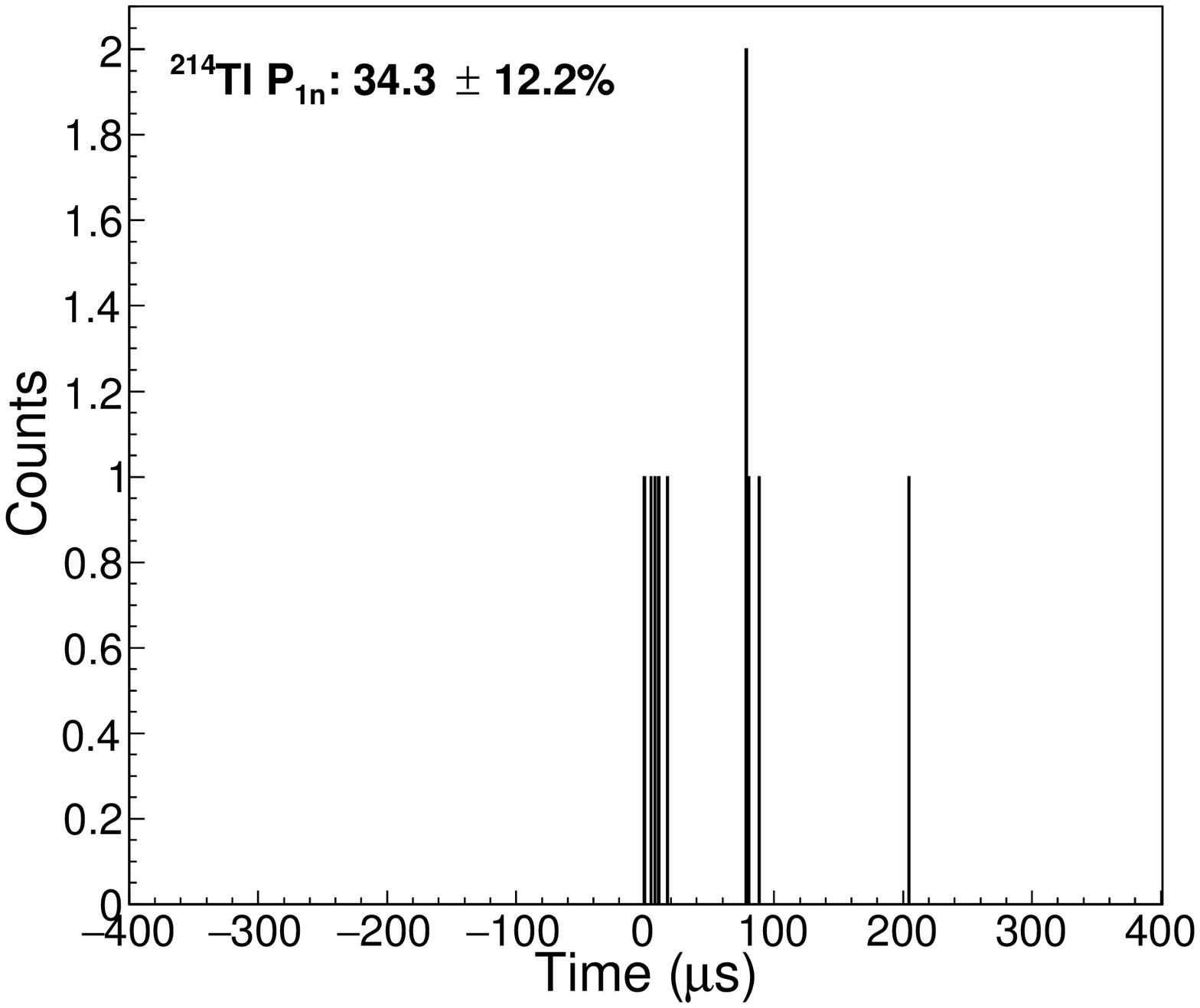}
  \caption{Implant-$\beta$ correlation histograms for $^{211,212,214,215,216}$Tl. Diagrams of $^{215,216}$Tl obtained using a correlation area of 5$\times$5~mm$^2$. Last histogram corresponds to the implant-$\beta$-neutron time-correlations for $^{214}$Tl.}
\label{F4-Tl}
\end{figure*}

Five more thallium isotopes were measured and their $\beta$-decay half-lives and neutron branching ratios were analyzed following the approach described above for $^{213}$Tl. The implant-$\beta$ time-correlation histograms for all of them, $^{211,212,214,215,216}$Tl are shown in Fig.~\ref{F4-Tl}. The binned ML-analysis (dashed-dotted line on diagrams) yields $T_{1/2}=76.5\pm17.8$~s for $^{211}$Tl, $T_{1/2}=30.9\pm8.0$~s for $^{212}$Tl and $T_{1/2}=11.0\pm2.4$~s for $^{214}$Tl. An almost negligible contribution from the much slower decay of the daughter nuclei ($^{211,212,214}$Pb) can be observed in these three cases. The most exotic thallium nuclei $^{215,216}$Tl were measured with rather limited statistics of only 281 and 99 implants, respectively. In this case, we noticed a slight improvement of the signal-to-background ratio in the correlation histograms when the correlation area was enlarged from $3 \times 3$~mm$^2$ to $5 \times 5$~mm$^2$. The ML analysis also shown in diagrams of Fig.~\ref{F4-Tl} yielded a half-life of $T_{1/2}=9.7\pm3.8$~s for $^{215}$Tl and $T_{1/2}=5.9\pm3.3$~s for $^{216}$Tl.

The neutron data analysis showed the presence of only one correlated implant-$\beta$-neutron event for each of $^{211,212,215}$Tl, which yields neutron branching ratios of 2.2(2.2)\%, 1.8(1.8)\% and 4.6(4.6)\%, respectively. For these three cases, given the low number of events compatible with the physical boundary, we have alternatively calculated a conservative upper-limit based on the Bayesian approach~\cite{gelman2014bayesian}, which yields upper limits of 10\%, 8\% and 20\% at a confidence level (CL) of 95\%. On the other hand, with BELEN we were able to observe a rather large number (10) of implant-$\beta$-neutron correlated events for $^{214}$Tl (histogram also shown in Fig.~\ref{F4-Tl}), resulting in a $P_{1n}$ value of $34.3\pm12.2~\%$. No single correlated or uncorrelated event was detected for $^{216}$Tl, which according to the implantation statistics obtained led an upper limit of 11.5\% and a Bayesian upper limit of $P_{1n}<52$\% at the 95\% CL. Concerning the other implanted species, $^{209,210}$Tl, the implantation statistics was not enough to determine either their half-lives nor the neutron branching ratios.

\subsection{Lead isotopes: $^{215-218}$Pb}
\begin{figure*}[!htbp]
\includegraphics[width = 0.38 \textwidth]{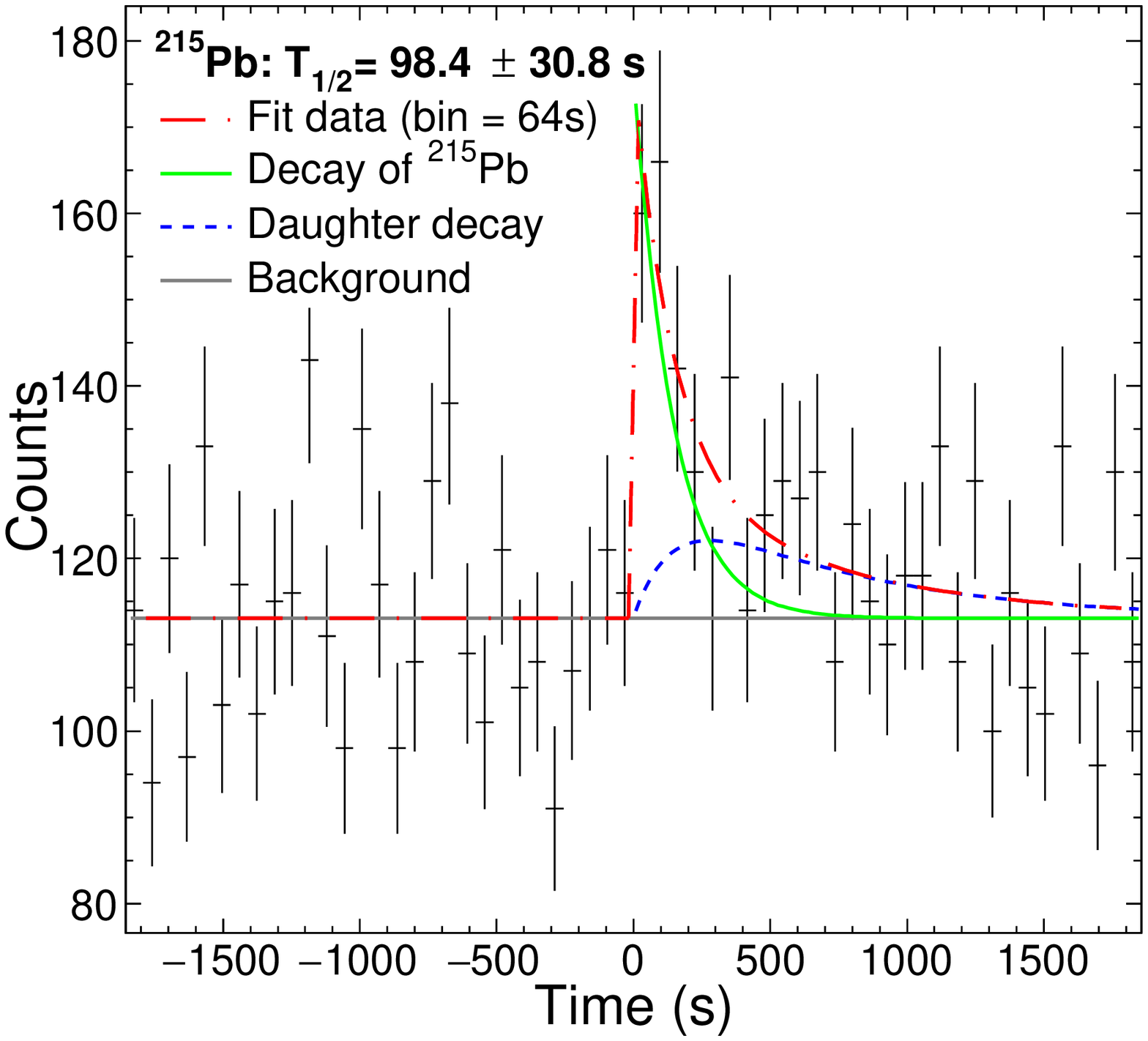}
\includegraphics[width = 0.38 \textwidth]{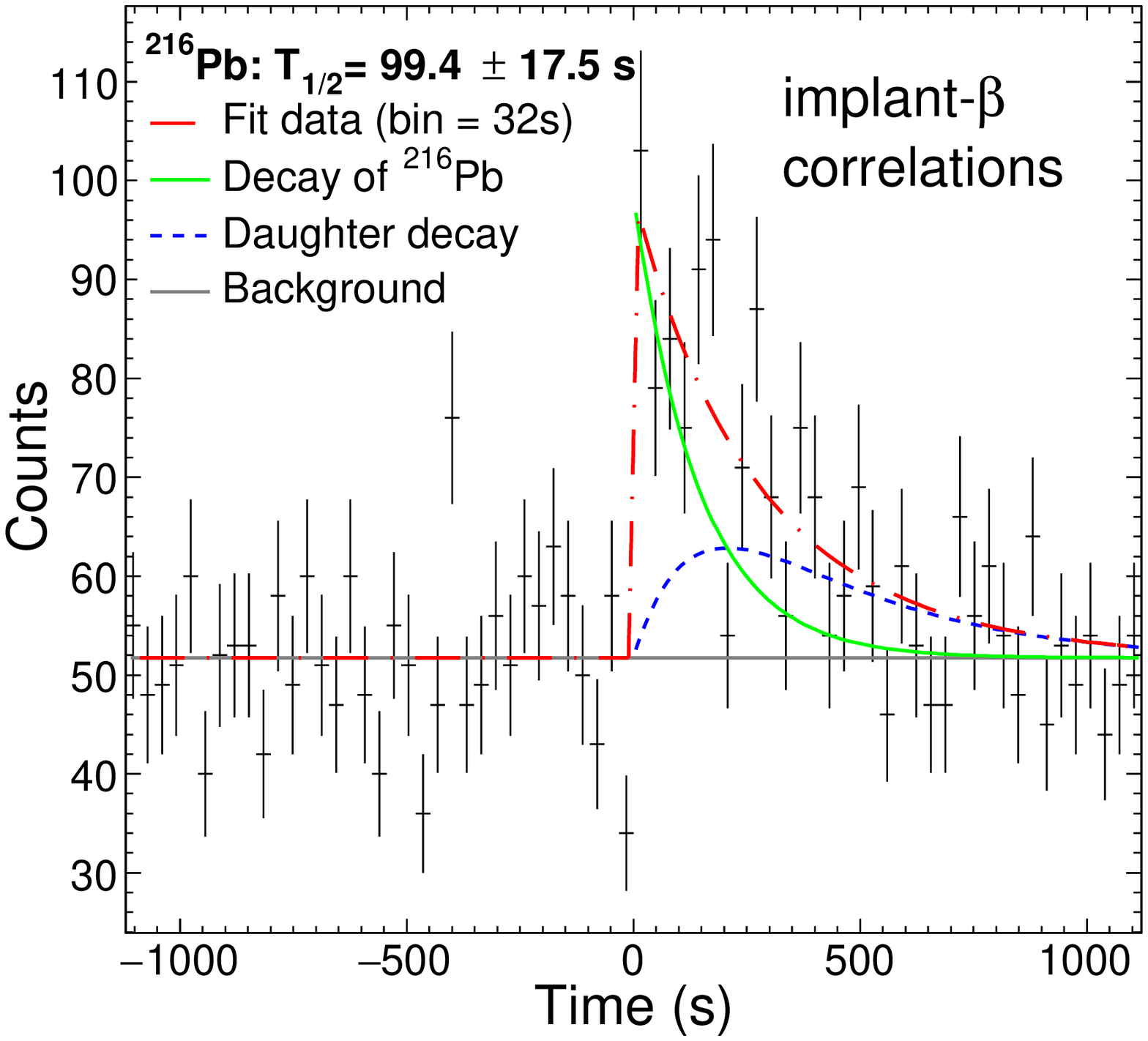}
\includegraphics[width = 0.38 \textwidth]{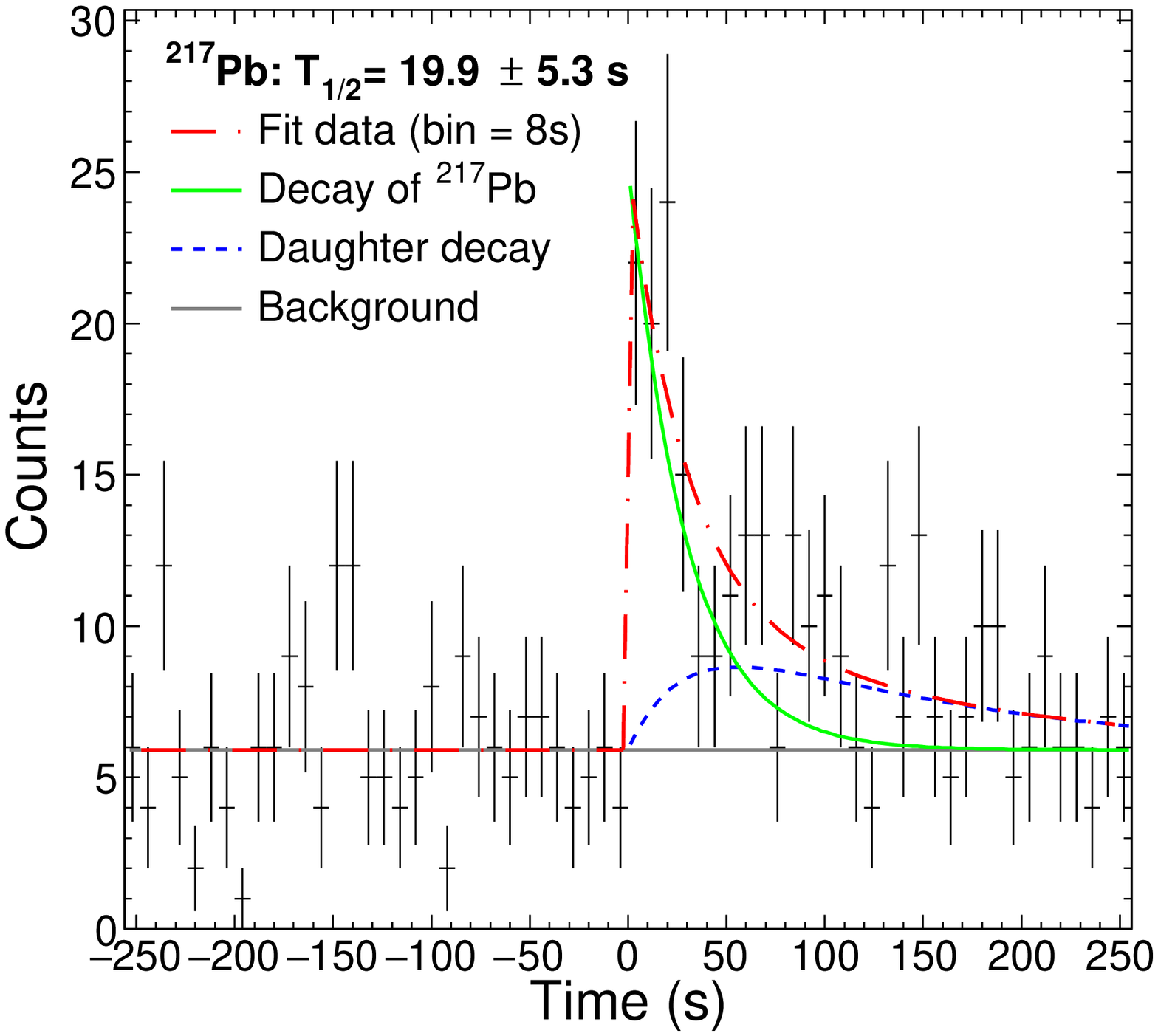}
\includegraphics[width = 0.38 \textwidth]{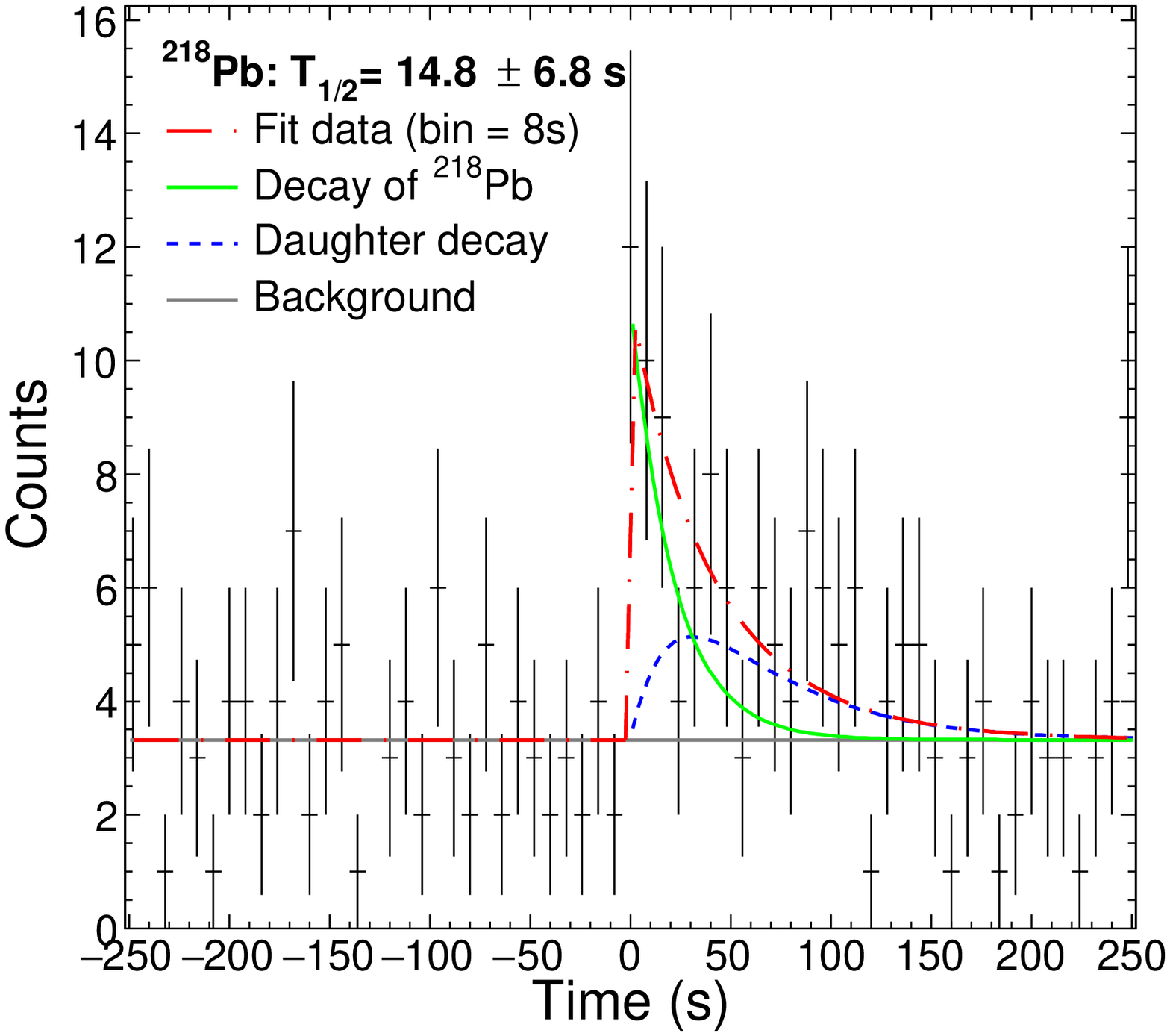}
\caption{Half-life analysis for implant-$\beta$ correlation diagrams of $^{215,216,217,218}$Pb. }
\label{F4-Pb}
\end{figure*}

\begin{figure}[!htbp]
\includegraphics[width = 0.78 \columnwidth]{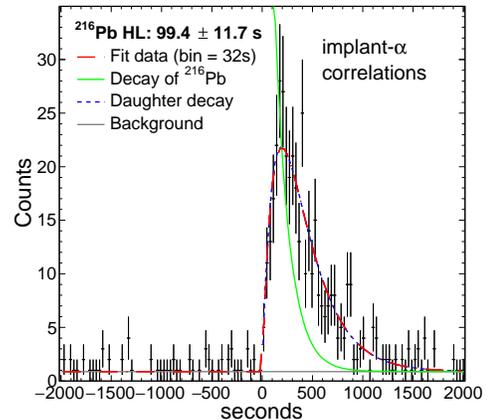}
  \caption{Half-life analysis of $^{216}$Pb via implant-$\alpha$ correlations.}
\label{F4-216Pb-alpha}
\end{figure}

The lead isotopes $^{212-219}$Pb were identified in the FRS and implanted in SIMBA. Out of them, $^{215-218}$Pb were implanted with enough statistics for a reliable half-life analysis. $^{214}$Pb was also implanted with large statistics (see Fig.~\ref{F3-NumImplants}), but its half-life of $1608\pm54$~s~\cite{nndcbnl} was too long for our analysis methodology and instrumentation. The ML analysis for the implant-$\beta$ correlation diagrams of $^{215-218}$Pb are shown in Fig.~\ref{F4-Pb}. 
The $^{215}$Pb analysis yielded a half-life of $T_{1/2}=98.4\pm30.7$~s, as it can be observed on its diagram, contributions from $^{215}$Pb and his daughter $^{215}$Bi are present in the correlation histogram. For the case of the measurement of $^{216}$Pb, it allowed us to carry out a cross-check of the aforementioned analysis methodology, as its half-life can be determined by means of two different methods: implant-$\beta$ and implant-$\alpha$ correlations. The former is also illustrated in Fig.~\ref{F4-Pb}, which shows the implant-$\beta$ correlation and the result of the ML analysis. On the other hand, the peak of $\alpha$-particles at 6778.5~keV, clearly identified in the energy spectrum of SIMBA (see Fig.~\ref{fig:dssd}b), corresponds to the decay of its granddaughter nucleus $^{216}$Po. Taking into account that the half-life of $^{216}$Po, 145$\pm$2~ms~\cite{wu2007nuclear}, is much shorter than that of the direct daughter, $^{216}$Bi, 2.25$\pm$5~s~\cite{wu2007nuclear}, it was possible to apply the method described in~\cite{morales2014beta} to obtain the half-life of $^{216}$Pb. With this method we determined a half-life of $T_{1/2}=99.4\pm11.7$~s (see diagram of Fig.~\ref{F4-216Pb-alpha}) which is in perfect agreement with the one obtained applying the conventional method described above for implant-$\beta$ time correlation, $T_{1/2}=99.4\pm17.5$~s. The accuracy in the analysis of $^{217}$Pb and $^{218}$Pb was mainly limited by the implantation statistics, which was of 436 and 235 implants, respectively. However, a reliable ML analysis was possible from their implant-$\beta$ correlation diagrams, as shown in the bottom diagrams of Fig.~\ref{F4-Pb}. The resulting half-lives were $T_{1/2}=19.9\pm5.3$~s for $^{217}$Pb and $T_{1/2}=14.8\pm6.8$~s for $^{218}$Pb. According the negative Q$_\beta n$-values of all these lead isotopes (see Table~\ref{T5-Pn}), no neutron branching emission is expected on them.

\subsection{Mercury isotopes: $^{208-211}$Hg}
From the eight mercury isotopes identified, $^{206-213}$Hg, implant statistics were high enough to analyze reliably four of them, $^{208-211}$Hg. The decay curve is strongly determined by the number of ion implants, as well as by the value of the half-life. 
\begin{figure*}[!htbp]
\includegraphics[width = 0.38 \textwidth]{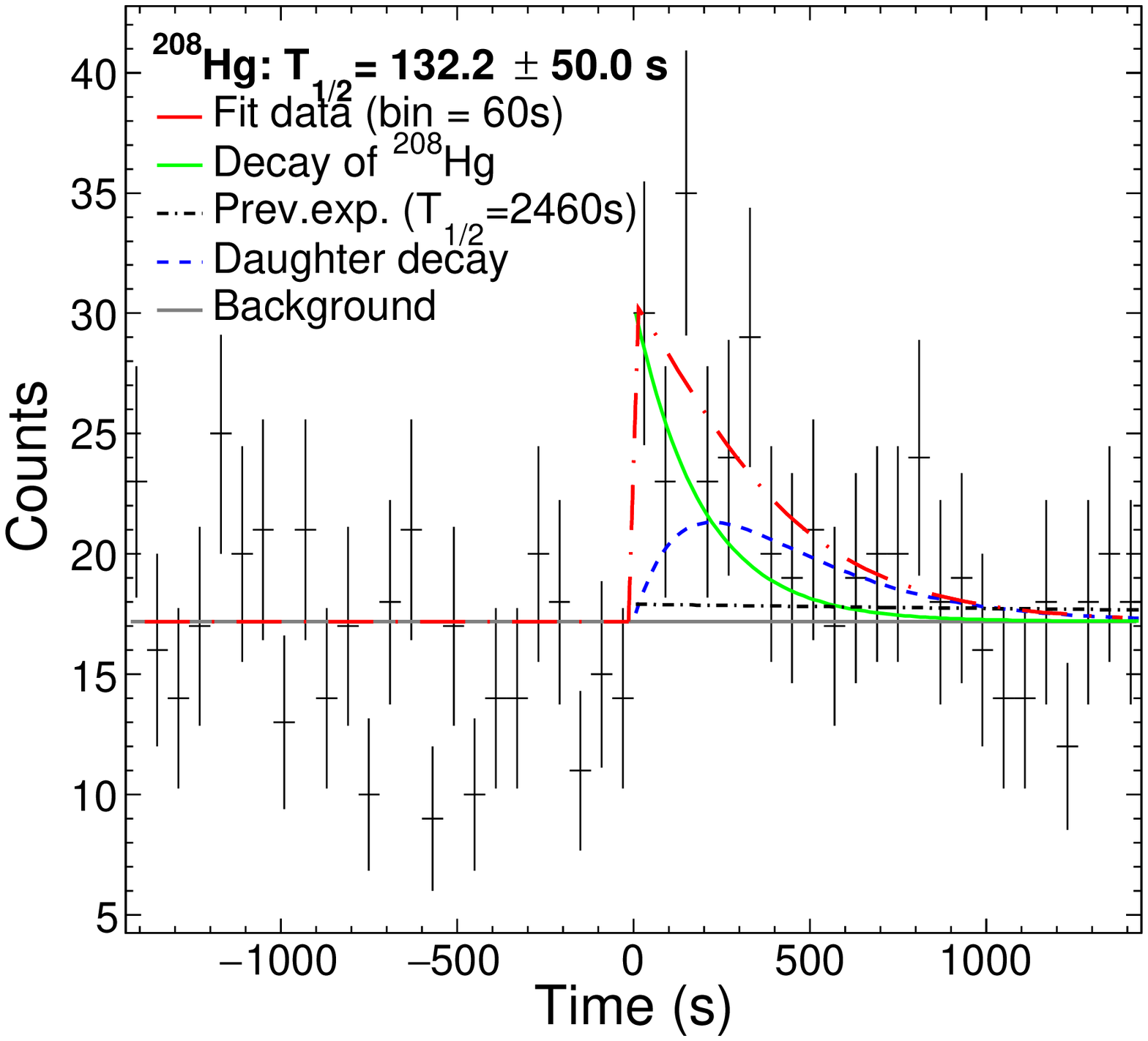}
\includegraphics[width = 0.38 \textwidth]{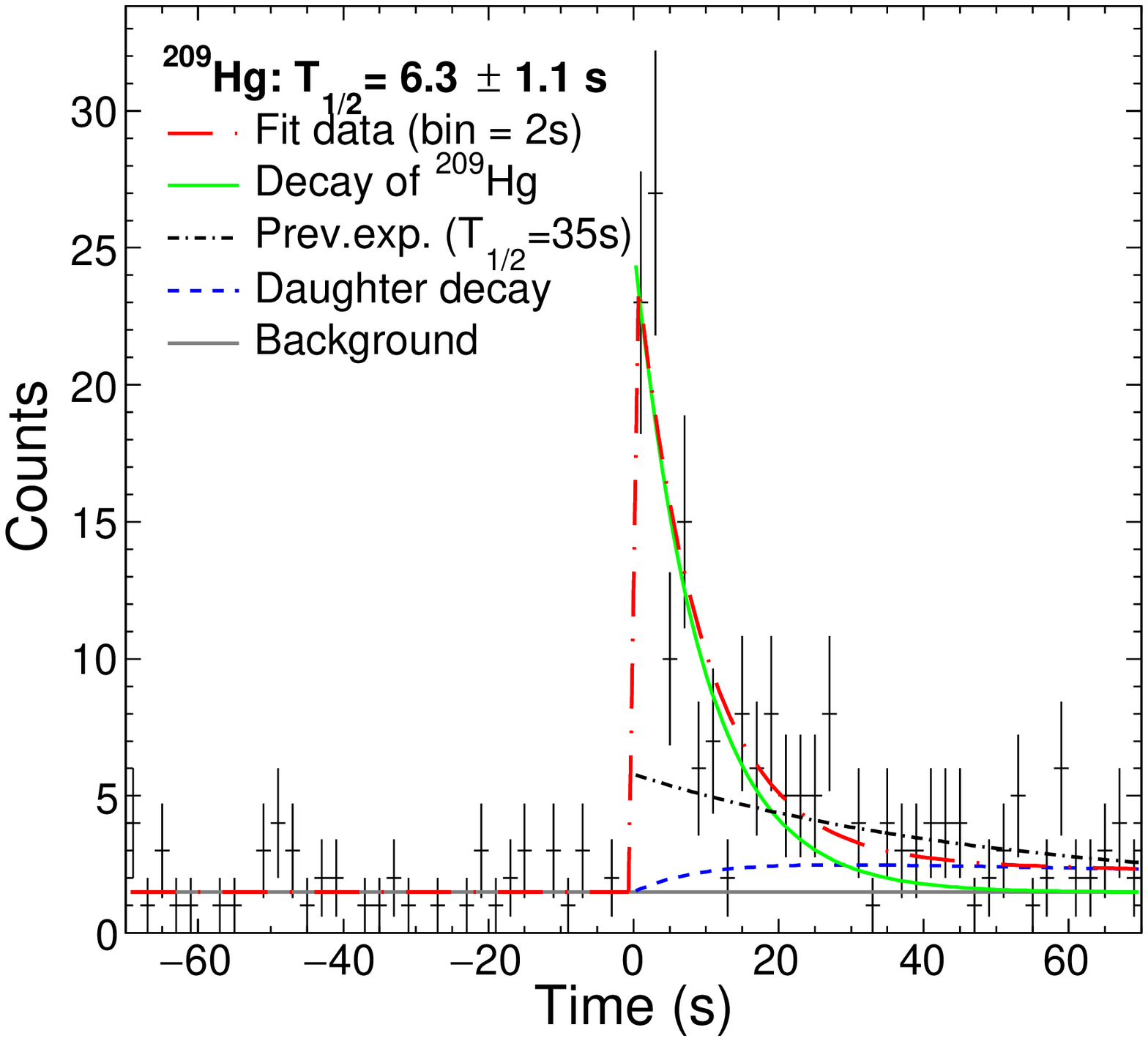}
\includegraphics[width = 0.38 \textwidth]{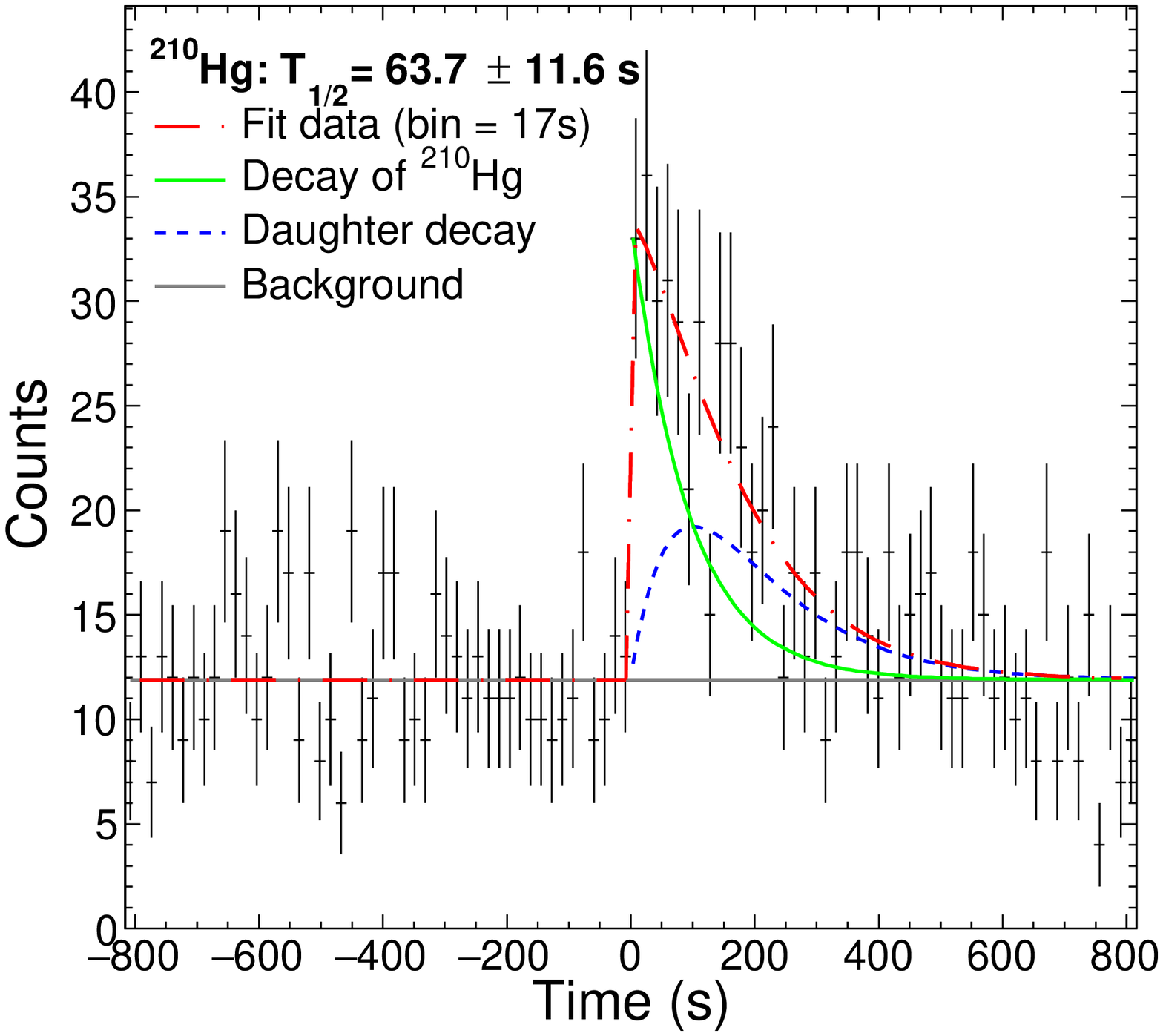}
\includegraphics[width = 0.38 \textwidth]{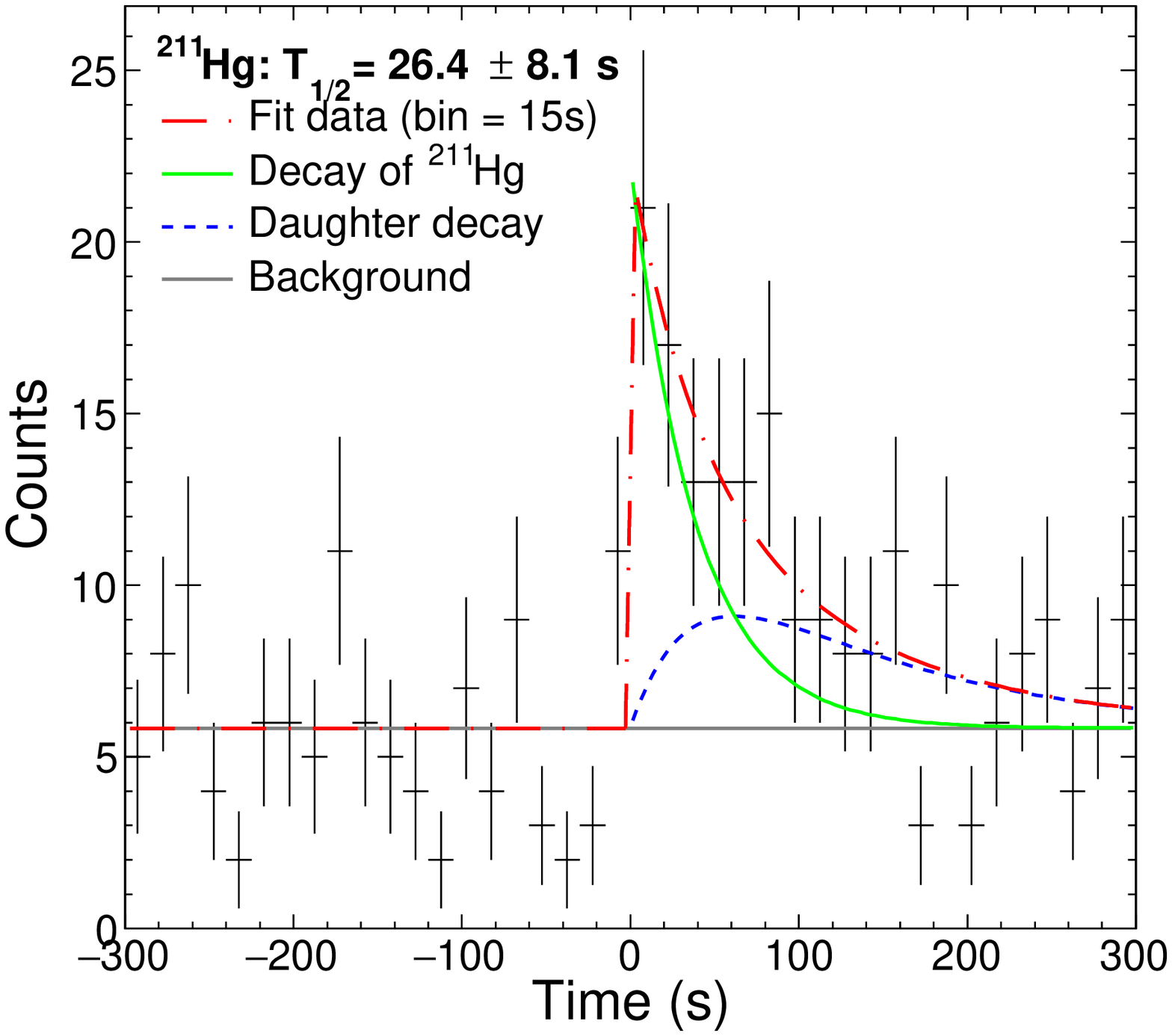}
\caption{Implant-$\beta$ correlation diagrams for $^{208,209,210,211}$Hg. In $^{208}$Hg and $^{209}$Hg diagrams is shown the parent and daughter contributions to the measured decay curve, and the decay of previous half-lives values (see text for details).}
\label{F4-Hg}
\end{figure*}
Thus, for $^{208}$Hg, with a low implantation statistics (220 events) and a relatively large half-life obtained from the ML analysis, $T_{1/2}=132.2\pm50$~s, the statistical uncertainty was 38\%. On the other hand, $^{209}$Hg has a factor of about two more implants (583 events) and a shorter half-life ($T_{1/2} = 6.3\pm1.1$~s), which leads to a much lower uncertainty of 17\%. An intermediate situation is found for the remaining two mercury isotopes, $^{210,211}$Hg, for which their half-lives analysis yielded $T_{1/2}=63.7\pm11.6$~s and $T_{1/2}=26.4\pm8.1$~s, respectively. In the latter case, the value obtained in the present work for the half-life of $^{211}$Tl was employed in the analysis. All aforementioned analyzed half-lives are shown in diagrams of Fig.~\ref{F4-Hg}. Regarding the neutron emission branching ratio, for $^{210}$Hg and $^{211}$Hg one implant-$\beta$-neutron event has been detected in the forward (moderation) time-window for each nucleus. This measurement yields neutron branching ratios of 2.2(2.2)\% and 6.3(6.3)\%, respectively. For these cases the calculated conservative upper-limit based on the Bayesian approach~\cite{gelman2014bayesian} yields upper constraints of 10\%, and 28\% at a CL of 95\%.

\subsection{Gold isotopes: $^{204-206}$Au}
The $^{203-209}$Au isotopes were identified in this measurement, but the implantation statistics was only high enough to analyze three of them, $^{204-206}$Au. 
\begin{figure}[!htbp]
\includegraphics[width = 0.78 \columnwidth]{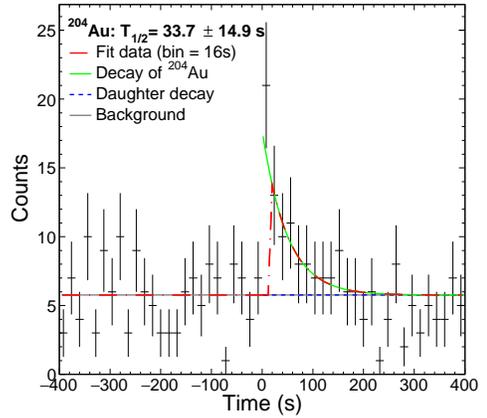}
\caption{$^{204}$Au implant-$\beta$ diagram. Correlation area of 25~mm$^2$ and extended including implant-$\beta$ events inside the spill.}
\label{F4-204Au}
\end{figure}
\begin{figure*}[!htbp]
\centering
\includegraphics[width = 0.38 \textwidth]{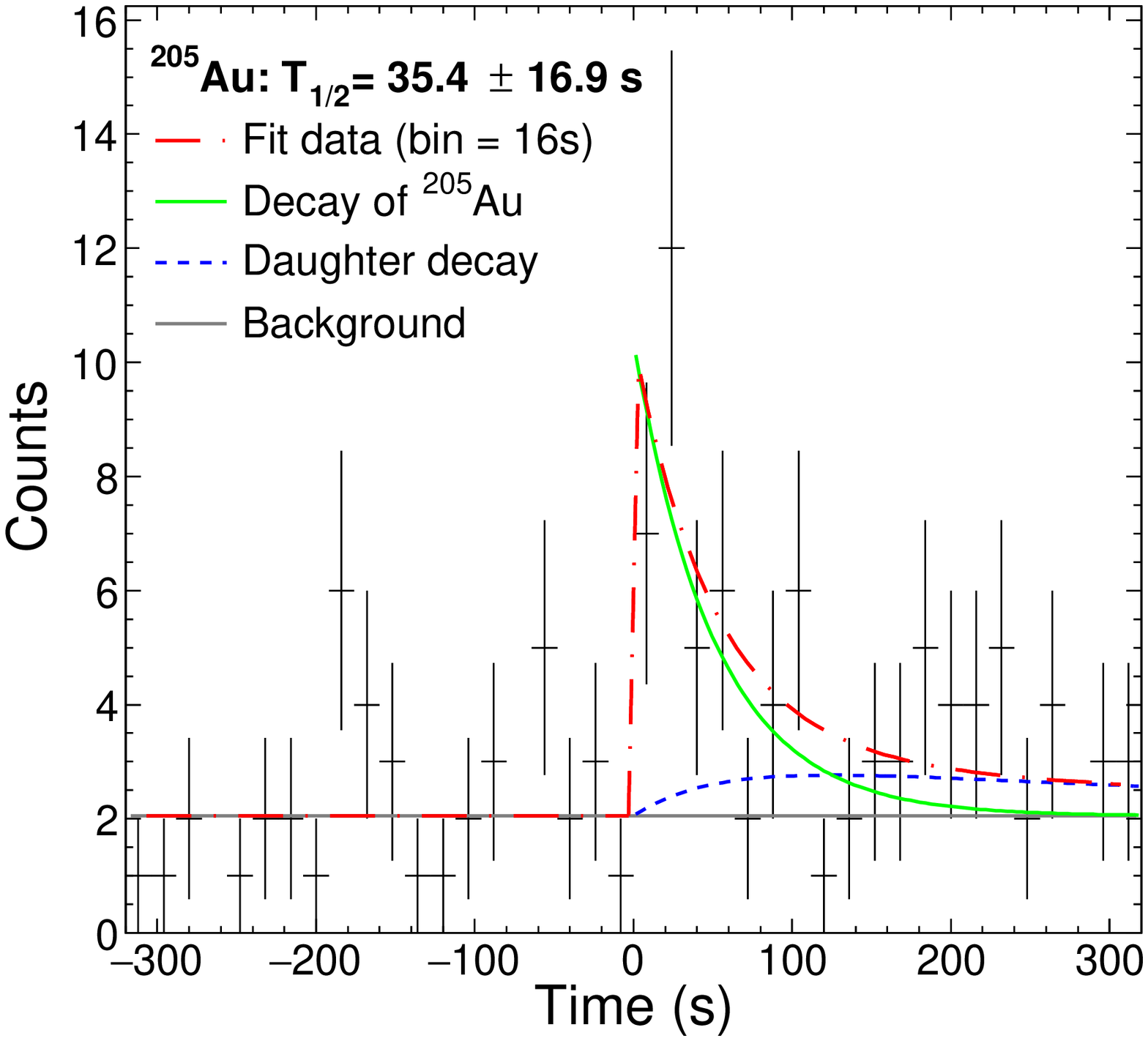}
\includegraphics[width = 0.38 \textwidth]{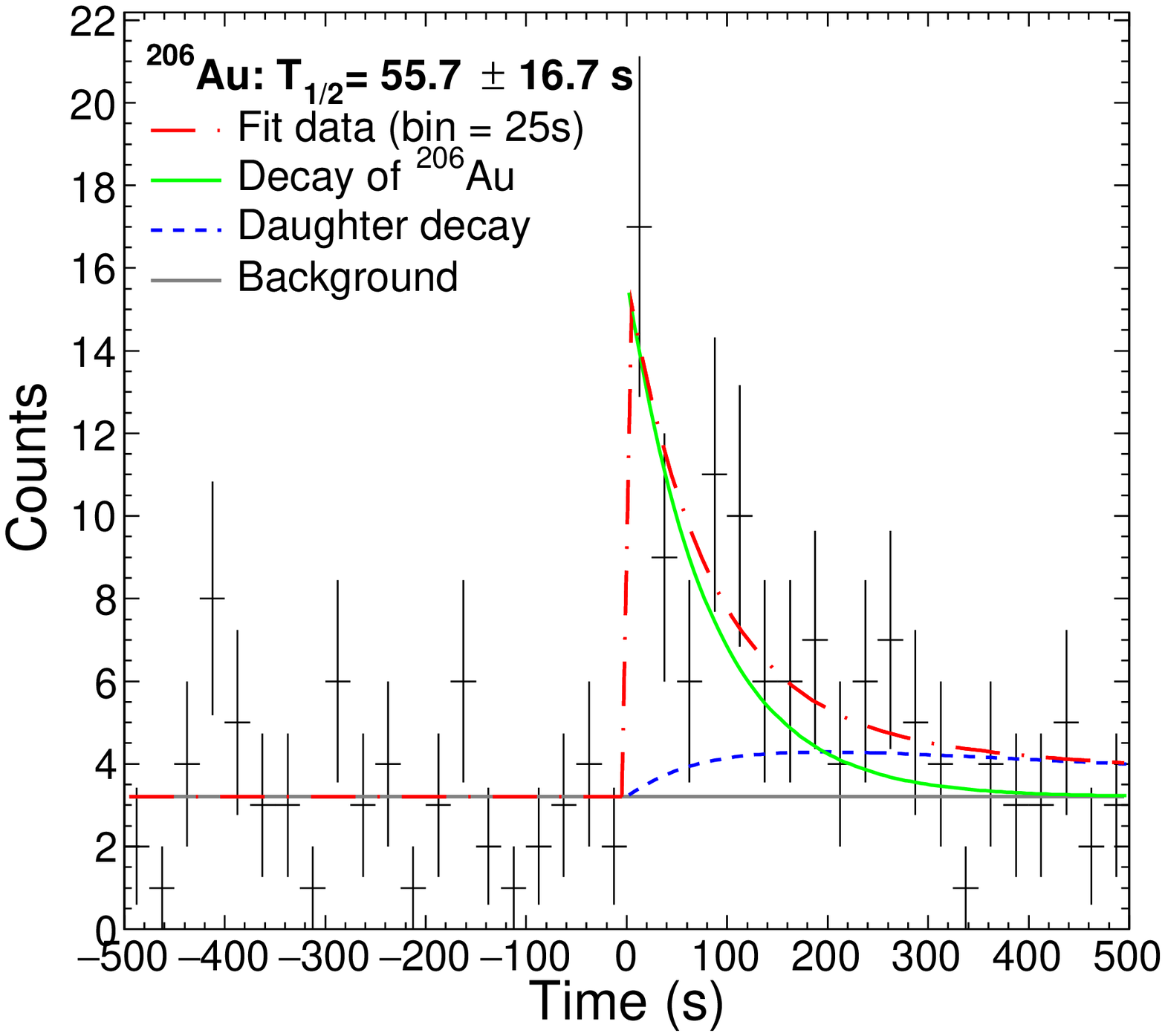}
\caption{Analysis of the implant-$\beta$ correlation diagrams for $^{205}$Au and $^{206}$Au.}
\label{F4-205-206Au}
\end{figure*}
In the case of $^{204}$Au (see Fig.~\ref{F4-204Au}) we also found a slight improvement of the decay curve when the correlation area was enlarged from 1~pixel (9~mm$^2$) to 2~pixels (25~mm$^2$) around the implant location, and included in the correlation those implant-$\beta$ events detected during the spill time. These provide a better sensitivity for the analysis, which yields a half-life of $T_{1/2}=33.7\pm14.9$~s. 
For the cases of $^{205}$Au and $^{206}$Au, both with $\sim 100$ implants, the ML analysis of the implant-$\beta$ correlation diagrams yield half-lives of $T_{1/2}=35.4\pm16.9$~s and $T_{1/2}=55.7\pm16.7$~s, respectively. Fig.~\ref{F4-205-206Au} shows the analysis of these two gold isotopes.

\subsection{Bismuth isotopes: $^{218-220}$Bi}\label{Results-Bi}
\begin{figure*}[!htbp]
\centering
\includegraphics[width = 0.38 \textwidth]{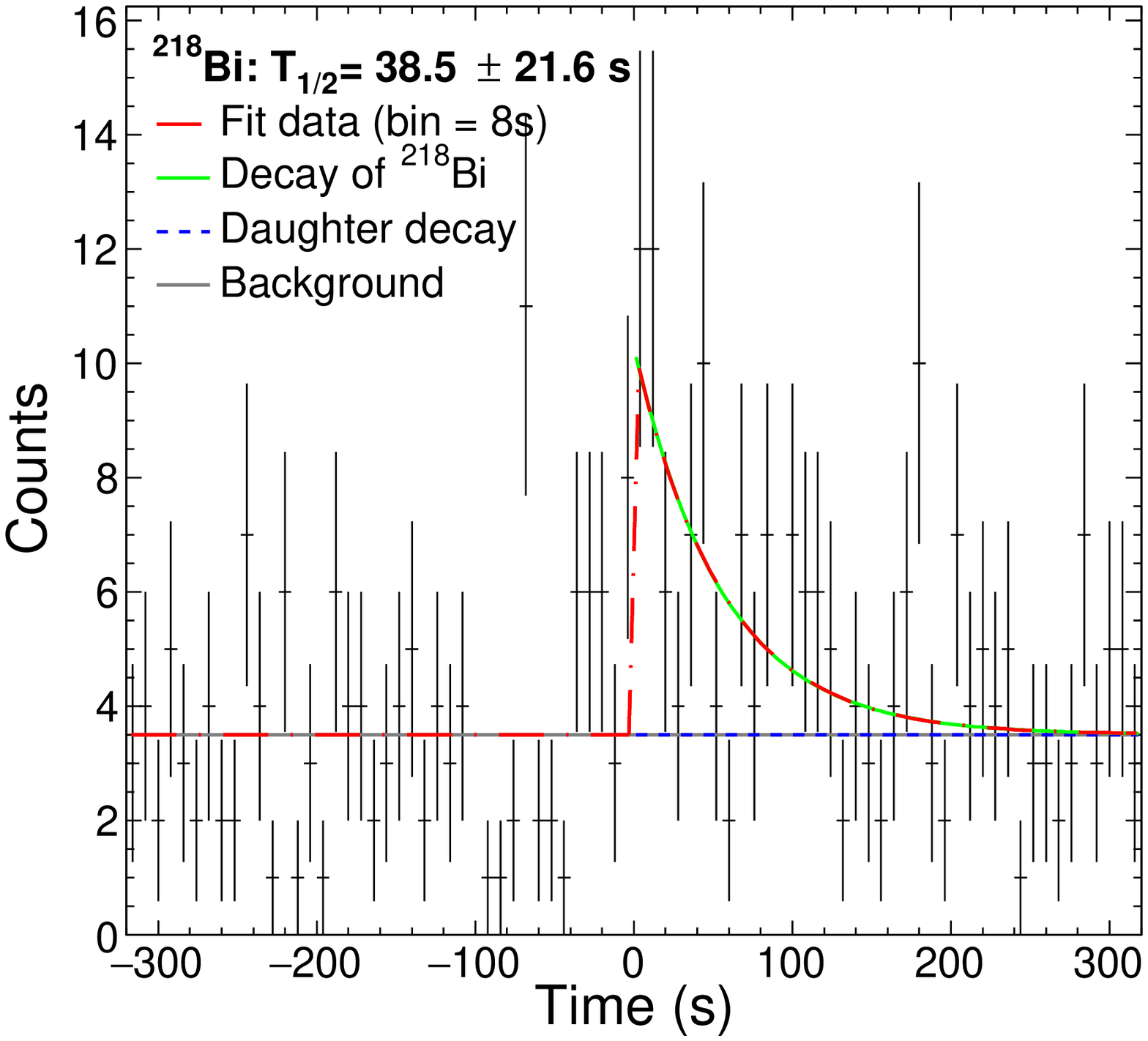}
\includegraphics[width = 0.38 \textwidth]{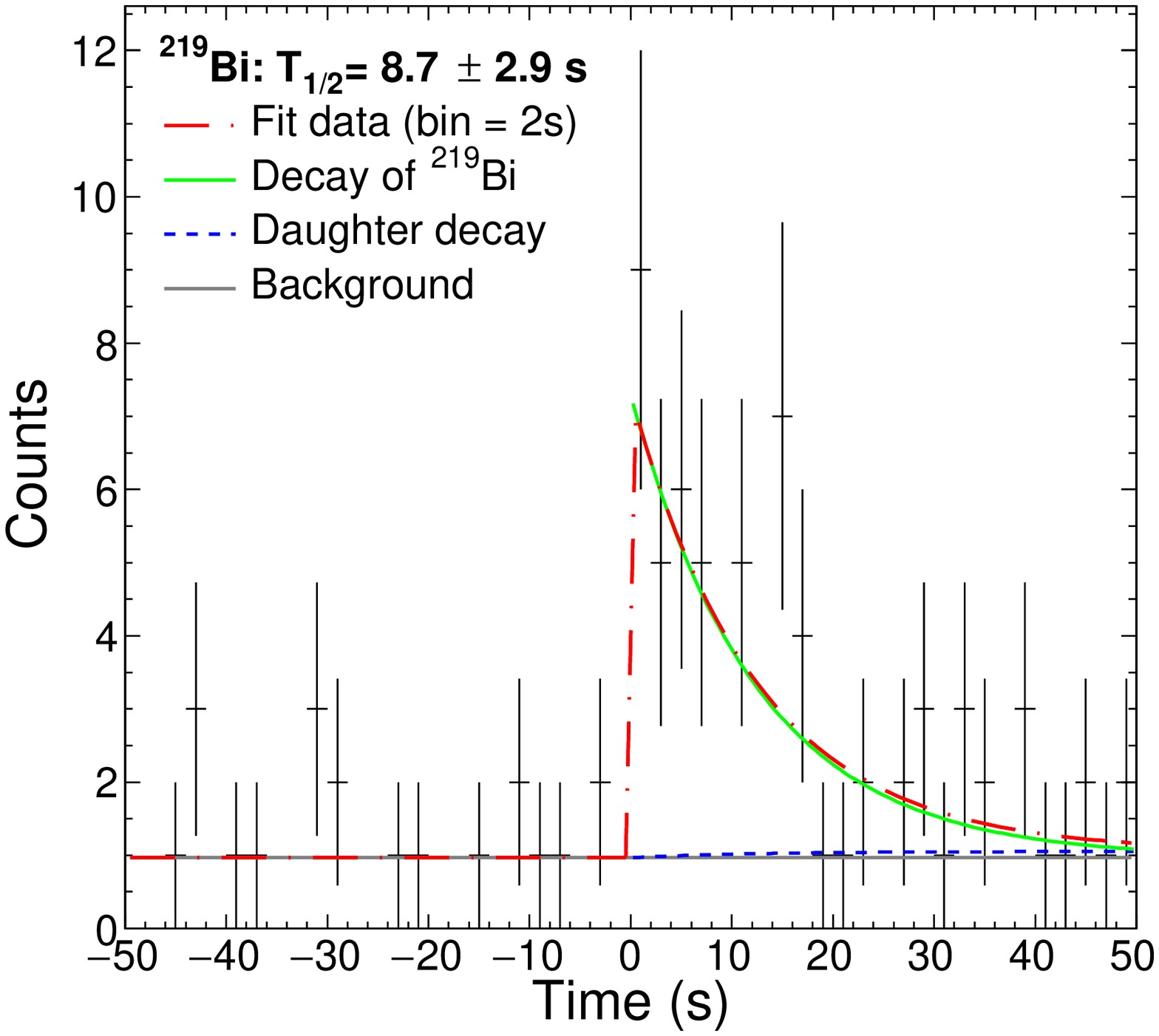}
  \caption{Analysis of the implant-$\beta$ correlation diagrams for $^{218}$Bi and $^{219}$Bi.}
\label{F4-218-219Bi}
\end{figure*}

Bismuth was the heaviest element implanted in SIMBA and we were able to determine the half-lives of three isotopes, $^{218-220}$Bi. The ML analysis of $^{218}$Bi yields a half-life of $T_{1/2}=38.5\pm21.6$~s and, as it can be seen in its diagram on Fig.~\ref{F4-218-219Bi}, the $\beta$ contribution comes only from its own decay, as its daughter ($^{218}$Pb) is an $\alpha$ emitter. The half-life analysis for $^{219}$Bi yields $T_{1/2}=8.7\pm2.9$~s (see the diagram also on Fig.~\ref{F4-218-219Bi}) and it includes the recent published half-life of its daughter, $^{219}$Po, $T_{1/2}=620\pm59$~s~\cite{fink2015source}.

\begin{figure}[!htbp]
\centering
\includegraphics[width = 0.78\columnwidth]{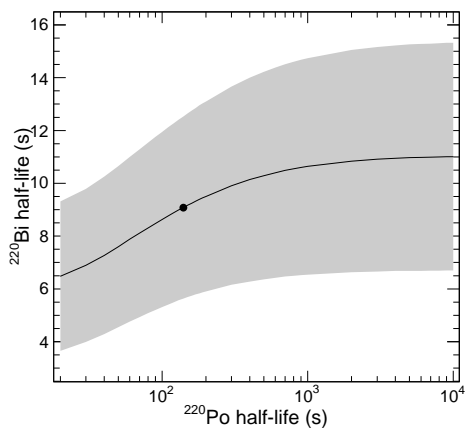}
  \caption{$^{220}$Bi half-life measured by varying the time of unknown daughter half-life in a wide range. (The dot indicates the resulting half-life using the FRDM+QRPA theoretical value for $^{220}$Po, $T_{1/2}=138.47$~s).}
\label{F4-220Bi}
\end{figure}

For the $\beta$-decay analysis of $^{220}$Bi one has to take into account that the half-life value of its daughter nucleus $^{220}$Po is still unknown. Thus, our analysis provides a range of possible half-life values for $^{220}$Bi, which spans between 4~s and 15~s. The bold marker in Fig.~\ref{F4-220Bi} represents the $^{220}$Bi half-life using the theoretical prediction calculated by FRDM+QRPA model~\cite{moller2003new} for the half-life of $^{220}$Po, $T_{1/2}=138.47$~s.

\section{Summary and discussion}\label{sec:discussion}
\begin{figure*}[!htbp]
\includegraphics[width = \textwidth]{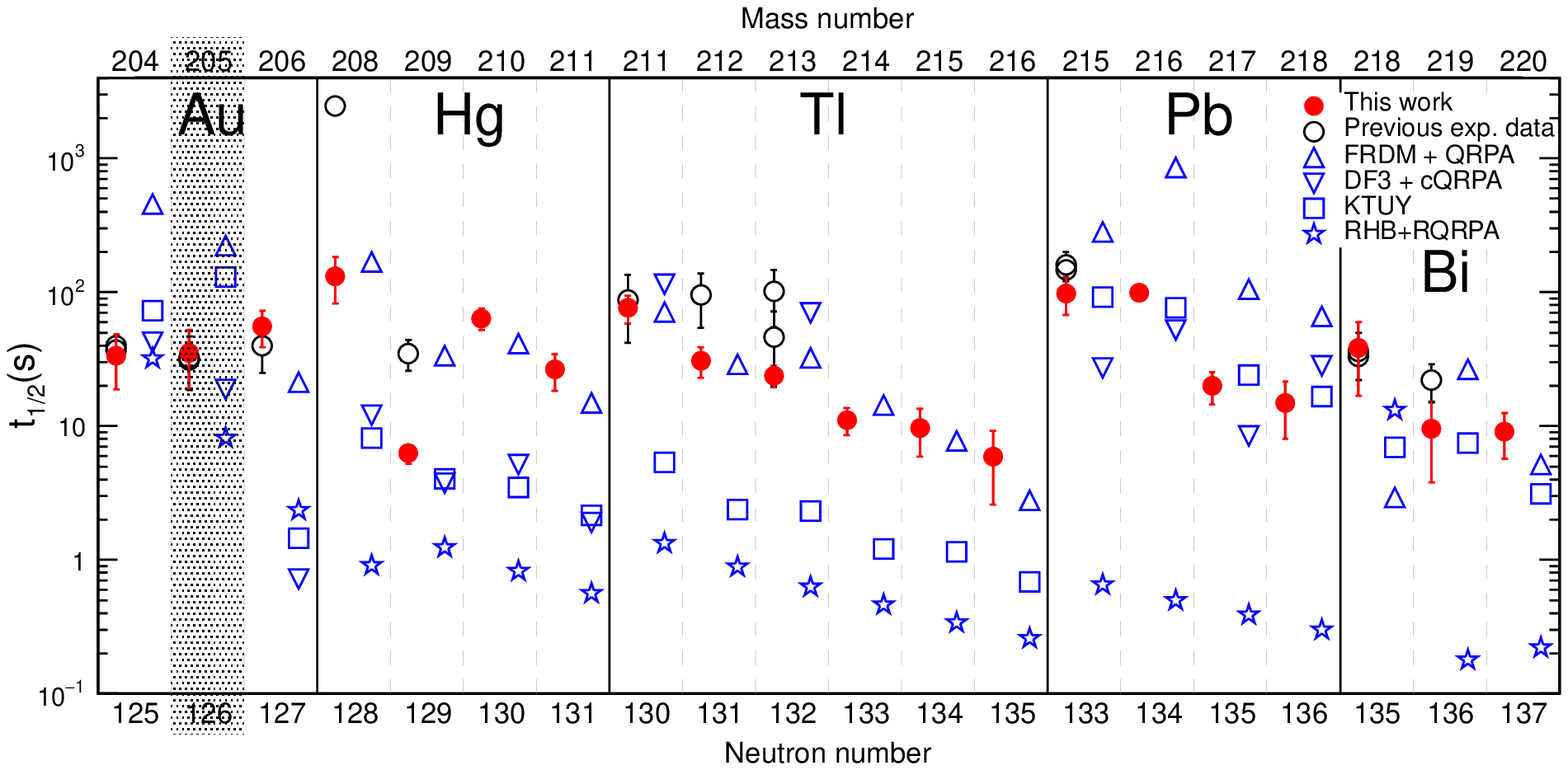}
\caption{Half-lives obtained in this work (red dots). Previous experimental values (open black circles). The blue symbols correspond to the theoretical values from different models: triangles for FRDM+QRPA~\cite{moller2003new} and DF3+cQRPA~\cite{borzov2006beta}, squares for the calculations from KTUY~\cite{koura2005nuclidic} and stars for the RHB+RQRPA~\cite{marketin2016large} model. See text and table~\ref{T5-HL} for details.}
\label{F5-ResultsHL}
\end{figure*}

\begin{figure}[!htbp]
\includegraphics[width = 1.0\columnwidth]{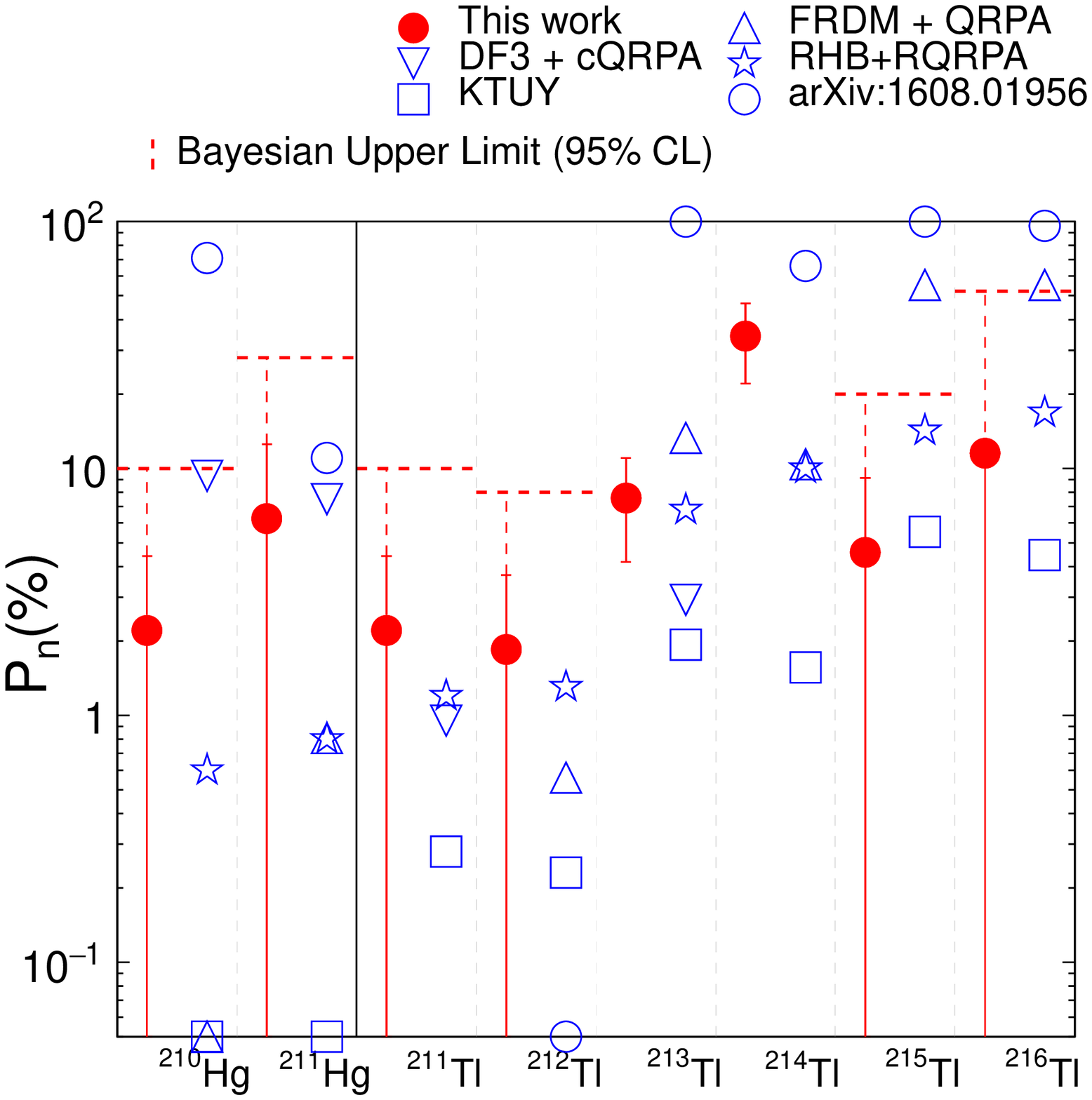}
\caption{Experimental neutron branching ratios and upper limits obtained in this work (red dots) and conservative upper-limit based on the Bayesian approach at a CL of 95\% (red dashed lines). The blue triangles correspond to the theoretical predictions by FRDM+QRPA~\cite{moller2003new} and DF3+cQRPA~\cite{borzov2006beta} models. Blue squares and stars to KTUY~\cite{koura2005nuclidic} and RHB+RQRPA~\cite{marketin2016large}, respectively, and blue circles to new recent calculations by~\cite{mumpower2016neutron}. See text and table~\ref{T5-Pn} for details.}
\label{F5-Pn}
\end{figure}

This section provides a summary of the main results obtained in this work, as well as a comparison with all previous experimental results and the theoretical predictions of  FRDM+QRPA~\cite{moller2003new} and DF3+cQRPA~\cite{borzov2006beta} models. A detailed comparison with the more recent calculations of Refs.~\cite{koura2005nuclidic, marketin2016large} can be found in Ref.\cite{Caballero-Folch2016First}. In summary, half-live values for 20 neutron-rich isotopes of Au, Hg, Tl, Pb and Bi have been determined experimentally, as well as neutron-branching ratios (or upper limits) for eight of them. As some of the analyzed isotopes are not expected to be neutron emitters according their Q$_{\beta n}$-values (see the last column of table~\ref{T5-Pn}), due to the low statistics available, the neutron emission analysis has been focused in those with large enough Q$_{\beta n}$. The results are displayed in Fig.~\ref{F5-ResultsHL} and Fig.~\ref{F5-Pn}, respectively. For comparison purposes, previously published theoretical and experimental half-life values have been also included in Fig.~\ref{F5-ResultsHL}.

Apart from the discrepancies found for the $^{208,209}$Hg isotopes, a rather good agreement is found between the present results and previously published experimental data. In the case of gold isotopes, all half-lives are fully compatible with those reported in previous measurements~\cite{craig204Au,morales2014half,wennemann1994investigation,morales2015first}, which are in reasonable agreement with DF3+cQRPA for $N\leq126$. For $^{206}$Au ($N=126$) the half-life obtained confirms the recent value published in Ref.~\cite{morales2015first} and both differ from the DF3+cQRPA value and the trends predicted by the available theoretical models. 
Whether this ``gold anomaly" is related to effects changing the occupation of the $\nu i_{11/2}$ orbital, or due to a weakening of the spin-orbit field caused by the tensor force~\cite{goddard2013charge}, or might be due to the recently proposed three-body force mechanism~\cite{nakada2015further,gottardo2012new} remains an open question which calls for more detailed theoretical studies and further specific experiments allowing for reconstruction of the decay scheme.

Concerning Hg isotopes, as discussed in Ref.~\citep{Caballero-Folch2016First}, recent measurements at CERN-ISOLDE~\cite{PrivCommZsolt2015} indicate that the half-life of $^{208,209}$Hg nuclei are much shorter than the values reported in Ref.~\cite{li1998neutron} and, therefore, this discrepancy will not be discussed further. Theoretical predictions by the FRDM+QRPA seem to agree better with the measured half-lives, with the only exception of $^{209}$Hg, whereas the opposite is found for the Hg half-lives predicted with DF3+cQRPA.

Regarding the thallium isotopes, previous half-lives were obtained in another experiment by using a similar experimental setup, but using a completely different analysis approach~\cite{benzoni2012first,kurtukian2008new}. In summary a good agreement is found for $^{211}$Tl, whereas the half-lives of $^{212}$Tl and $^{213}$Tl differ by factors of 2-3. It is worth emphasizing the overall good agreement for the thallium chain between the present results and FRDM+QRPA predictions, including the case of the most exotic nuclei reported here for the first time, $^{214-216}$Tl. This result seems to indicate the rather low relevance of first forbidden (FF) transitions in the $N\geq126$ mass region, as discussed in Ref.~\cite{Caballero-Folch2016First}.

The analysis of the lead isotopes includes three new half-lives, $^{216-218}$Pb. The half-life obtained for $^{215}$Pb is in reasonable agreement with the two previous measurements~\cite{sagawa1987nuclear,morales2014beta}. Along the lead isotopic chain, the FRDM+QRPA model overestimates the experimental values by factors between 3-9. The values predicted by DF3+cQRPA are, on average, slightly closer to the measured half-lives, but the mass (neutron number) dependency of the half-life is not satisfactorily reproduced by any of these two models.

For the bismuth chain the FRDM+QRPA predictions agree reasonably well with the heaviest measured nuclei $^{219-220}$Bi, whereas almost one order of magnitude difference is found for $^{218}$Bi. The latter half-life is, however, rather well established experimentally~\cite{benzoni2012first}. The result quoted in this work for the half-life of $^{220}$Bi, 4-15~s (see Fig.~\ref{F4-220Bi}), can be re-determined more accurately once the half-life of $^{220}$Po is measured.

The neutron-branching ratios determined in this work represent the first set of experimental data available in this mass region. Therefore, the values reported here can be only compared with theoretical predictions. Both theoretical and experimental $P_{1n}$-values are listed in Table~\ref{T5-Pn} and displayed in Fig.~\ref{F5-Pn}.

In summary, the agreement between theory and experiment is rather good for the Hg and Tl isotopes with masses between 208 and 214. However, for the two heaviest thallium nuclei $^{215,216}$Tl, KTUY~\cite{koura2005nuclidic} and RHB+RQRPA~\cite{marketin2016large} models are in agreement but the other theoretical predictions of the neutron emission overestimate substantially the obtained experimental results.

\section{Conclusions}\label{sec:conclusions}
Since both the half-life and the neutron emission probability are integral quantities of the $\beta$-decay, it is difficult to explain why the FRDM+QRPA reproduces so well the average decay strength over the full Q$_{\beta}$-window (see Fig.~\ref{F5-ResultsHL}) along the chain of measured Tl isotopes, whereas it seems to fail dramatically in the upper energy range of $^{215,216}$Tl, beyond the neutron separation energy of the daughter nuclei, as deducted from their determined neutron branching ratios. At first sight, one is tempted to attribute such a discrepancy to the possible contribution of high-energy first forbidden (FF) transitions populating low-lying levels of the daughter nuclei and hindering therefore the emission of neutrons. Nevertheless, this interpretation is at variance with the overall systematic found in the $N\geq126$ region~\cite{morales2014beta,benzoni2012first,morales2014half,morales2015first,Caballero-Folch2016First}, where Gamow-Teller (GT) transitions seem to play a dominant role in general. At present, the only plausible explanation for such a feature could be that the overall strength of the $\beta$ decay is overestimated in FRDM+QRPA both, in the full Q$_{\beta}$ and in the upper energy window Q$_{\beta n}$. As reported in Ref.~\cite{Caballero-Folch2016First}, more advanced microscopic models such as RHB+RQRPA~\cite{marketin2016large} and KTUY~\cite{koura2005nuclidic} show an inverted behavior, yielding good predictions for the $P_{1n}$-values, but discrepant values for the half-lives, thus not improving the situation. Recent calculations based on an improved QRPA and HF theory~\cite{mumpower2016neutron}, included in Fig.~\ref{F5-Pn}, also show large discrepancies. Clearly more $\beta$-decay measurements and theoretical efforts are needed in this mass region, in order to gain a better understanding of the underlying nuclear structure effects, as well as to guide global theoretical models far-off stability.

\begin{acknowledgments}
This work was supported by the \textit{Spanish Ministerio de Economia y Competitividad} under grants No. FPA2011-28770-C03-03, FPA2008-04972-C03-03, AIC-D-2011-0705, FPA2011-24553, FPA2008-6419, FPA2010-17142, FPA2014-52823-C2-1-P, FPA2014-52823-C2-2-P, CPAN CSD-2007-00042 (Ingenio2010), and the program Severo Ochoa (SEV-2014-0398). I.D and M.M were supported by the German Helmholtz Association via the Young Investigators Grant VH-NG 627 (LISA- Lifetime Spectroscopy for Astrophysics) and the Nuclear Astrophysics Virtual Institute (VH-VI-417), and by the \textit{German Bundesministerium f\"ur Bildung und Forschung} under No. 06MT7178 / 05P12WOFNF. R.C.F acknowledges the support of the Spanish Nuclear Security Council (CSN) under a grant of Catedra Argos. UK authors acknowledge the support of the UK Science \& Technology Facilities Council (STFC) under grant No. ST/F012012/1. Yu.A.L. acknowledges support from Helmholtz-CAS Joint Research Group (HCJRG-108). R.C.F. and I.D. are also supported by the National Research Council of Canada (NSERC) Discovery Grants SAPIN-2014-00028 and RGPAS 462257-2014 at TRIUMF.
\end{acknowledgments}

\begin{table*}[ht!]
\caption{Half-lives ($T_{1/2}$) results, previous experimental data and theoretical predictions.\label{T5-HL}}
\begin{ruledtabular}
\begin{tabular}{cccccccc}
\\
Nuclei&$N$& Implanted& T$^{exp}_{1/2}$~(s)& Previous& FRDM+QRPA& DF3+cQRPA\\
&& ions&(this work)&T$^{exp}_{1/2}$~(s)&(s)~\cite{moller2003new}&(s)~\cite{borzov2006beta}\\
\hline
\\
$^{204}$Au& 125& ~~54& 33.7~$\pm$14.9& 39.8~$\pm$0.9~\cite{craig204Au}& 455.3& 42.4\\
&&&&37.2~$\pm$0.8~\cite{morales2014half}&&\\
$^{205}$Au& 126& ~103& 35.4~$\pm$16.7& 31.0~$\pm$0.2~\cite{wennemann1994investigation}& 222.0& 18.7\\
&&&&32.5~$\pm$14.0~\cite{morales2014half}&&\\
$^{206}$Au& 127& ~106& 55.7~$\pm$16.7& 40.0~$\pm$15.0~\cite{morales2015first}&21.3& 0.72\\
\hline
\\
$^{208}$Hg& 128& ~220& 132.2~$\pm$50.0& 2460$^{+300}_{-240}$~\cite{li1998neutron}& 168.9& 12.1\\
$^{209}$Hg& 129& ~583& ~6.3~$\pm$1.1& 35.0$^{+9}_{-6}$~\cite{li1998neutron}& 33.6& 3.7\\
$^{210}$Hg& 130& ~512& 63.7~$\pm$11.6& $>$300~ns& 41.2& 5.2\\
$^{211}$Hg& 131& ~253& 26.4~$\pm$8.1& $>$300~ns& 14.9& 1.9\\
\hline
\\
$^{211}$Tl& 130& ~483& 76.5~$\pm$17.8& 88$_{-29}^{+46}$~\cite{benzoni2012first}& 70.9& 114.9\\
$^{212}$Tl& 131& 1056& 30.9~$\pm$8.0& 96$_{-38}^{+42}$~\cite{benzoni2012first}& 29.0& -\\
$^{213}$Tl& 132& 1015& 23.8~$\pm$4.4& 101$_{-46}^{+486}$~\cite{chen2010discovery}& 32.4& 70.4\\
&&&&46$_{-26}^{+55}$~\cite{benzoni2012first}&&\\
$^{214}$Tl& 133& ~598& 11.0~$\pm$2.4& $>$300~ns& 14.4& -\\
$^{215}$Tl& 134& ~281& 9.7~$\pm$3.8& $>$300~ns& 7.8& -\\
$^{216}$Tl& 135& ~~99& 5.9~$\pm$3.3& $>$300~ns& 2.8& -\\
\hline
\\
$^{215}$Pb& 133& 1079& 98.4~$\pm$30.8& 147~$\pm$12~\cite{sagawa1987nuclear}& 282.5& 27.1\\
&&&&160~$\pm$40~\cite{morales2014beta}&&\\
$^{216}$Pb& 134& 1005& 99.4~$\pm$11.7& $>$300~ns& 852.2& 52.0\\
$^{217}$Pb& 135& ~436& 19.9~$\pm$5.3& $>$300~ns& 104.9& 8.5\\
$^{218}$Pb& 136& ~235& 14.8~$\pm$6.8& $>$300~ns& 66.3& 28.4\\
\hline
\\
$^{218}$Bi& 135& ~294& 38.5~$\pm$21.6& 33~$\pm$1~\cite{nndcbnl}& 2.92& -\\
&&&&36~$\pm$14~\cite{benzoni2012first}&&\\
$^{219}$Bi& 136& ~306& 8.7~$\pm$2.9& 22~$\pm$7~\cite{benzoni2012first}& 26.54& -\\
$^{220}$Bi& 137& ~176& 9.5~$\pm$5.7 & $>$300~ns& 5.17& -\\
\end{tabular}
\end{ruledtabular}
\end{table*}

\begin{table*}[ht!]
  \caption{$P_{n}$ results compared with theoretical predictions of measured isotopes.\label{T5-Pn}}
  \begin{ruledtabular}
  \begin{tabular}{ccccccccccc}
\\
Nuclei&  $N$& $P_{n}$(\%)&FRDM+QRPA& DF3+cQRPA&RHB+RQRPA&KTUY&arXiv:1608.01956& $Q_{\beta n}$(keV)~\cite{audi2012TheAmeI,audi2012TheAmeII}\\
&&(this work)&(\%)~\cite{moller2003new}&(\%)~\cite{PrivCommBorzov2010}&(\%)~\cite{marketin2016large}&(\%)~\cite{koura2005nuclidic}&(\%)~\cite{mumpower2016neutron}& (extr. = extrapolated)\\
\hline
\\
$^{204}$Au& 125& - & 0.0& 0.0&0.1&0.0&-& $-$3453~$\pm$200 (extr.)\\
$^{205}$Au& 126& - & 0.0& 0.0&0.2&0.0&-& $-$2151~$\pm$196 (extr.)\\
$^{206}$Au& 127& - & 0.0& 0.0&0.5&0.0&0.0& 2~$\pm$298 (extr.)\\
\\
$^{208}$Hg& 128& - & 0.0& 3.2&0.3&0.0&-& $-$303.32~$\pm$31.23\\
$^{209}$Hg& 129& - & 0.0& 2.8&0.5&0.0&-& 34~$\pm$149 (extr.)\\
$^{210}$Hg& 130& 2.2~$\pm$2.2& 0.0& 9.3&0.6&0.0&71& 201~$\pm$196 (extr.)\\
$^{211}$Hg& 131& 6.3~$\pm$6.3& 0.81& 7.5&0.8&0.0&11& 551~$\pm$196 (extr.)\\
\\
$^{211}$Tl& 130& 2.2~$\pm$2.2& 0.04& 0.95&1.2&0.28&-& 578.67~$\pm$41.95\\
$^{212}$Tl& 131& 1.8~$\pm$1.8& 0.56& - &1.3&0.23&0.0& 869~$\pm$200 (extr.)\\
$^{213}$Tl& 132& 7.6~$\pm$3.4& 13.26& 2.93&6.8&1.93&100& 1259.73~$\pm$27.10\\
$^{214}$Tl& 133& 34.3~$\pm$12.2& 10.38& - &10&1.56&66& 1595~$\pm$196 (extr.)\\
$^{215}$Tl& 134& 4.6~$\pm$4.6& 55.24& - &14.2&5.54&100& 2021~$\pm$298 (extr.)\\
$^{216}$Tl& 135& $<$11.5&55.36& - &17&4.45&96& 2230~$\pm$315 (extr.)\\
\\
$^{215}$Pb& 133& - & 0.0& 0.0&0.3&0.0&-& $-$2455~$\pm$102 (extr.)\\
$^{216}$Pb& 134& - & 0.0& 0.0&0.3&0.0&0.0& $-$2240~$\pm$196 (extr.)\\
$^{217}$Pb& 135& - & 0.0& 0.0&0.4&0.0&0.0& $-$1705~$\pm$298 (extr.)\\
$^{218}$Pb& 136& - & 0.0& 0.0&0.4&0.0&0.0& $-$1348~$\pm$299 (extr.)\\
\\
$^{218}$Bi& 135& - & 0.0& - &1.2&0.0&0.0& $-$740.61~$\pm$27.73\\
$^{219}$Bi& 136& - & 0.06& - &0.3&0.0&0.0& $-$148~$\pm$196 (extr.).\\
$^{220}$Bi& 137& - & 0.01& - &0.4&0.0&0.0& 66.0~$\pm$298 (extr.)\\
\end{tabular}
\end{ruledtabular}
\end{table*}

%

\end{document}